\documentclass[journal,one column]{IEEEtran}
\usepackage{amsfonts, amsmath, amssymb, amsthm}
\usepackage{verbatim}
\usepackage{cite}

\usepackage{tikz}
\usetikzlibrary{arrows,automata}
\tikzstyle{state}=[circle,fill=!50,text=white,minimum size =25pt,inner sep=0pt]
\tikzstyle{selected state}=[state, fill=!100]
\tikzstyle{Markov state}=[circle,fill=!50,text=white,minimum size =40pt,inner sep=0pt]
\usepackage{url}
\usepackage{epstopdf}
\usepackage{epsfig}
\begin{document}

\newtheorem{Lemma}{Lemma}
\newtheorem{Theorem}{Theorem}
\newtheorem{Example}{Example}
\newtheorem{defn}{Definition}
\newtheorem{problemstatement}{Problem Statement}
\newtheorem{corollary}{Corollary}
\newtheorem{claim}{Claim}
\newtheorem{fact}{Fact}
\newtheorem{property}{Property}
\newtheorem{rmk}{Remark}

\def \hence {\Rightarrow}
\def\Sec {\S}
\newcommand \remove[1] {}
\newcommand \hide[1] {}
\newcommand \green[1]         {{\color{green}#1}}
\newcommand \blue[1]         {{\color{blue}#1}}
\newcommand \yellow[1]         {{\color{yellow}#1}}
\newcommand \red[1]         {{\color{red}#1}}

\newcommand{\bA}{A_{\epsilon,\delta}}
\newcommand{\bP}{{\psi (n, p, \delta,\epsilon)}}

\title{Pricing for profit in Internet of Things \thanks{A part of this paper was presented in ISIT\rq{}15\cite{iot_isit}.}}
\author{Arnob Ghosh, \and Saswati Sarkar \thanks{The authors are with the Electrical and Systems Engg. Department of University of Pennsylvania, Philadelphia, PA. Their e-mail ids are--arnob@seas.upenn.edu and swati@seas.upenn.edu.}}
\maketitle
\begin{abstract}
The economic model of the Internet of Things (IoT) consists of end users, advertisers and three different kinds of providers--IoT service provider (IoTSP), Wireless service provider (WSP) and cloud service provider (CSP). We investigate three different kinds of interactions among the providers. First, we consider that the IoTSP prices a bundled service to the end-users, and the WSP and CSP pay the IoTSP (push model). Next, we consider the model where the end-users independently pay the each provider (pull model). Finally, we consider a hybrid model of the above two where the IoTSP and WSP quote their prices to the end-users, but the CSP quotes its price to the IoTSP. 

We model different kinds of interaction among the providers as a combination of sequential and parallel non-cooperative games. We characterize and quantify the impact of the advertisement revenue on the equilibrium pricing strategy and payoff of providers, and corresponding demands of end users  in each of the above interaction models.  Our analysis reveals that the demand of end-users, and the payoffs of the providers are non decreasing functions of the advertisement revenue. For sufficiently high advertisement revenue, the IoTSP will offer its service free of cost in each interaction model. However, the payoffs of the providers, and the demand of end-users vary across different interaction models.
Our analysis shows that the demand of end-users, and the payoff of the WSP are the highest  in the pull (push, resp.) model in the low (high, resp.) advertisement revenue regime. The payoff of the IoTSP is always higher in the pull model irrespective of the advertisement revenue. The payoff of the CSP is the highest in the hybrid model in the low advertisement revenue regime. However, in the high advertisement revenue regime the payoff of the CSP in the hybrid model or in the push model can be higher depending on the equilibrium chosen in the push model. 
\end{abstract}
\vspace{-0.1cm}
\section{Introduction}
\subsection{Motivation}
The Internet of Things (IoT) is a world-wide network where various kinds of objects such as sensors, RFID tags, robots and smart phones will interact and operate with minimal human intervention.
 The IoT has several applications-- such as in smart grid, e-Health service,  smart transportation system,  building, and home automation. Research has been initiated to combat the technical challenges associated with the IoT \cite{survey,middleware}, however, the  economic aspects of the IoT remain largely ignored. Large scale proliferation of the IoT will critically depend on rendering it profitable to all the entities in the IoT paradigm. We seek to contribute in this space.

\subsection{Different Entities and modes of interactions}
We first characterize the entities involved in an IoT architecture: end-users (e.g. individuals, organizations, government agencies), advertisers, and providers. There are three different kinds of providers in an IoT-- the IoT Service Provider (IoTSP), the Wireless Service Provider (WSP), and the Cloud Service Provider (CSP). An IoT service will be provided by the IoTSP (e.g. Apple \footnote{Apple is going to launch a smart home-kit platform for smart home appliances very soon.}, General Electric) similar to Internet service provided by the Internet service provider (ISP). The  CSP (e.g. Amazon EC2) will help to store and process the enormous amount of data generated in the IoT. The WSP (e.g. At\&T) will enable the communication amongst devices, and between the devices and the providers\footnote{Since there will be a large number of devices in the IoT, thus, the communication will be mainly through wireless media. Thus, we consider a WSP, but our analysis will go through for wired service providers.}.   Online advertisement will generate revenue for the IoTSP as well as additional traffic for the WSP and CSP, and additional reimbursement for them\footnote{In the IoT, customized advertisement will be popular depending on the IoT service. For example, for smart vehicle system advertisement of auto-insurance can be seen.}. 

\begin{figure}
\includegraphics[width=90mm, height=60mm]{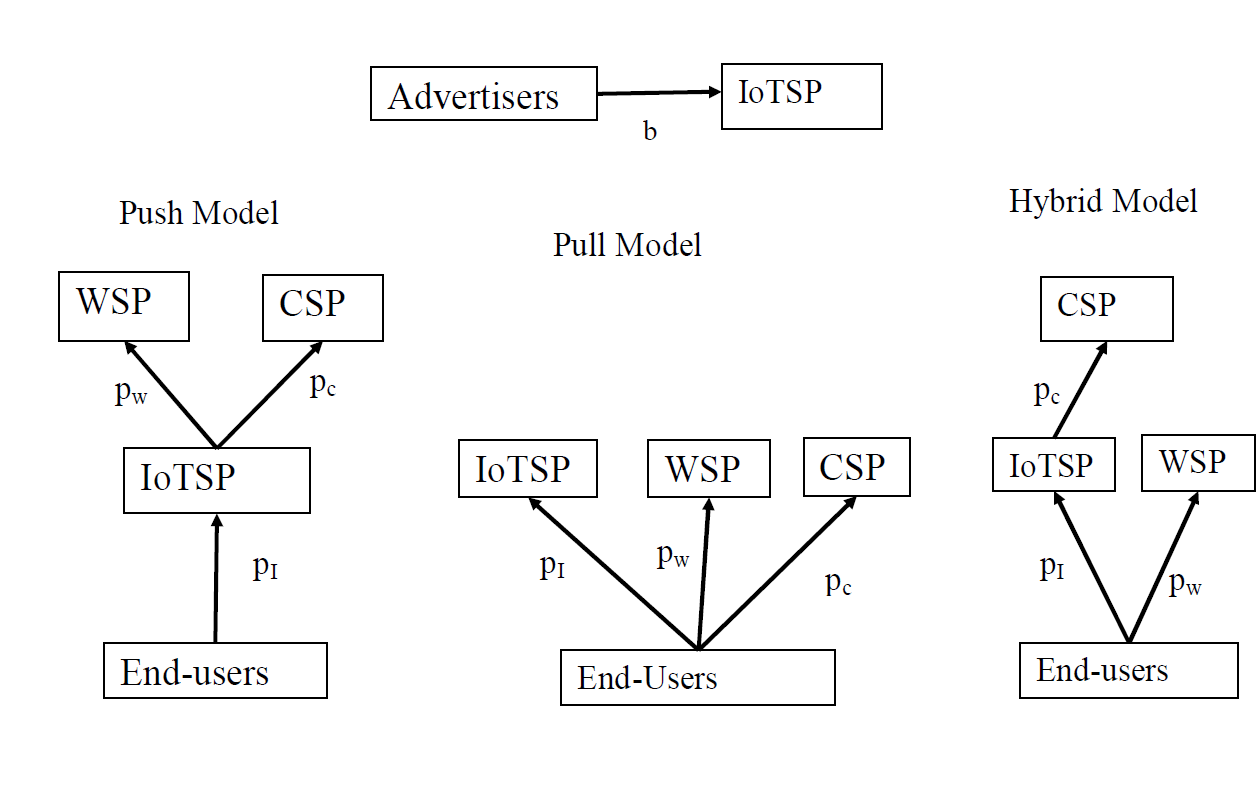}
\caption{Different interaction models: Push framework is depicted in the left hand side figure, pull and a hybrid frameworks are depicted in the center and the right hand side respectively. In each interaction model, advertisers pay the IoTSP. Directed arrows denote the direction in which money flows.}
\label{fig:push_pull_hybrid}
\vspace{-0.5cm}
\end{figure}

   The IoT market is still not widely deployed\footnote{AT\&T is offering smart home security system where it offers a bundled service to the end-users. However, the market is still in very early stage.}. When it will be widely deployed, the interaction among the entities is expected to be similar to the legacy technologies such as the internet and app markets. Towards this end, we investigate three different interaction models which are analogous to the existing models in legacy technologies. In the {\em push} or centralized model, the IoTSP procures the bandwidth and computing resources from the WSP and the CSP respectively and \lq\lq{}pushes" the bundled service  at a price $p_I$ to end-users  (Fig.~\ref{fig:push_pull_hybrid}). The term \lq\lq{}push\rq\rq{} is drawn from the  literature on the pricing in the Internet and Telecommunications where the \lq\lq{}push\rq\rq{} corresponds to the setting when a service provider pushes the data to the end-user \cite{push_pull1,push_pull2}.  An example of the interaction model analogous to the push model is  the Kindle service by Amazon.  Amazon (analogous to the IoTSP) charges only the end-users while delivering the  e-books to the Kindle devices using the wireless service of Whispernet (analogous to the WSP); Whispernet only gets its revenue from the Amazon. 

On the other extreme in a decentralized setting or {\em pull} model, each provider separately quotes a price to end-users, and an end-user \lq\lq{}pulls"  the service from the providers by paying them separately (Fig. \ref{fig:push_pull_hybrid}). Similar to the push, the name \lq\lq{}pull" is again drawn from the literature on the pricing in the Internet and Telecommunication where the \lq\lq{}pull" corresponds to the setting  when an end-user \lq\lq{}pulls\rq\rq{} the data from the service provider \cite{push_pull1,push_pull2}. Interaction models similar to the pull model are also seen in practice. For example,  both the ISP (analogous to the WSP) and some content providers (e.g. Netflix, analogous to the IoTSP) charge the end-users for the content.   

We also consider a hybrid of the above two models, where the IoTSP and WSP separately quote their prices to the end-users, but the CSP quotes its price to the IoTSP only (Fig.~\ref{fig:push_pull_hybrid}). The analogous interaction model is also prevalent in the app market. For example, in the current app markets (e.g. Google play store) end-users pay the WSP and the application providers (analogous to the IoTSP) for their services. End-users do not pay the CSP. The application providers pay the CSP (e.g. Google cloud service). 

\subsection{Technical Challenges and Objectives}
We now discuss the technical challenges involved in analyzing different interaction models. Towards this end, we first discuss the similarities and dissimilarities among different models.  In all these models, the demand of end-users decreases (increase, respectively) with the increase (decreases, resp.) in price that an end-user has to pay. In the pull model, the prices charged by all providers directly influence the demand. In the push (hybrid, resp.) model, the price charged by the IoTSP (IoTSP and WSP, resp.) {\em directly} influences the demand. On the other hand the price charged by the WSP, and  the CSP (the CSP, resp.) {\em indirectly}  influence the demand in the push (hybrid, resp.) model. The indirect influence in the push (hybrid, resp.) model is the following-- if the WSP and the CSP (the CSP, resp.) increase their prices charged to the IoTSP, then the IoTSP also has to increase its price quoted to end-users which in turn will decrease the demand. Thus, the providers\rq{} payoffs are not maximized at the extreme points.  We characterize pricing strategies of  providers in different interaction models. 

We now discuss the impact of the advertisement revenue on each provider and the demand of end-users. Advertisement revenue {\em directly} benefits the IoTSP and {\em indirectly} benefits the WSP and the CSP.  The indirect influence is the following-- in the push (pull or hybrid, resp.) model the IoTSP (end-user, resp.) has to pay the WSP  for the additional traffic which increases the WSP\rq{}s payoffs. Additional computing resources are also required for advertisement, thus, end-users (IoTSP, resp.) have to pay the CSP  in the pull model (push or hybrid, resp.) which increases the CSP\rq{}s payoff.  The advertisement revenue might benefit the end-users. This is because advertisement revenue may enable the IoTSP to decrease its price  leading to an increase in demand. However, the IoTSP (end-users, resp.) also has to pay the WSP more for the additional advertisement traffic in the push (pull or hybrid, resp.)  model which decreases the payoff of the IoTSP (the demand of end-users, resp.).  Thus, the impact of the advertisement revenue on the demand of the end-users and the payoffs of the providers is also not apriori clear.

With our analysis we seek to characterize the answers of the following fundamental questions:
 \begin{itemize}
 \item How do the IoTSP, WSP and CSP select their prices to the end-users or amongst each other depending on the interaction model? How does the IoTSP quote its price to the advertisers?
 \item How will the advertisement revenue impact the demand of the end-users (or, the reach of IoT service) and the pricing strategies of the providers in different interaction models? How does the advertisement revenue will impact the payoffs of the providers?
 \end{itemize}
 The answers of the above questions will provide us an insight on the followings--
 \begin{itemize}
 \item Among different interaction models, in which model the demand of end-users (or, the reach of the IoT service) will be the highest?
 \item Which model will provide the highest payoff to each provider and advertisers?
 \end{itemize}
 
 The above questions are of substantial significance.   When the IoT-market will be widely deployed, it is expected to be one of the two types-- i) a free market  where each of the above interaction models can co-exist, ii) a regulated market where the interaction model can be chosen by a policy maker. If the market is free,  a provider (either the IoTSP, WSP or CSP)  may want to adopt the interaction model which will fetch him the highest payoff.  Thus, the provider may want to know which interaction model will fetch him the highest payoff.    On the other hand, if the market is regulated, then the policy maker wants to adopt an interaction model which satisfies its objective.    For example, if the goal is to enhance the reach of the IoT technology, the policy maker would prefer to adopt the interaction model where the demand will be the highest. On the other hand, if the policy maker wants to incentivize  a provider (either IoTSP, WSP or the CSP), then it will adopt the interaction model that will be preferable to the provider. Thus,  the policy maker needs to quantify the payoffs of the entities in each interaction model. 
\subsection{Our Contributions}
\subsubsection{Problem Formulation}
We develop an economic framework that models three different interaction models described above. We consider a monopolistic setting i.e. there is one IoTSP, one WSP and one CSP. We consider a non-cooperative game between the advertisers and the IoTSP where advertisers decide to participate in the advertisement depending on the price set by the IoTSP and the end-users\rq{} preference (Section~\ref{game:IoTSP and advertisers}). Then, the IoTSP, the WSP and the CSP select their prices in different interaction models (push, pull and hybrid). We formulate a non-cooperative game theoretic setting where the IoTSP, WSP and CSP select their respective  prices in order to maximize the individual payoff in each of the three interaction models.

We model the price selection problem in the push interaction as a combination of a {\em sequential } and {\em parallel} non cooperative game. In the first stage of the sequential game, the WSP and the CSP quote their prices in parallel, in the second stage, the IoTSP selects its  price with the knowledge of the prices quoted by the WSP and the CSP (Section~\ref{sec:push_model}).  We characterize the sub-game perfect Nash equilibrium (SPNE) in the above game. In the pull model, since an end-user independently pays the providers,  we consider a non-cooperative game where providers select prices for end-users in parallel (Section~\ref{sec:pull_model}).  We characterize the Nash equilibrium in the above game. In the hybrid model, since the IoTSP pays the CSP and end-users independently pay the WSP and the IoTSP,  we consider a combination of sequential and parallel games, where in the first stage of the game the CSP quotes its price to the IoTSP and then in the last stage, the IoTSP and the WSP select prices in parallel (Section~\ref{sec:hybrid}). We characterize the SPNE in the above. 

The characterization of the equilibrium pricing strategies will provide us an insight on the questions posed above. The natural question is whether there exists a unique or multiple equilibria pricing strategies. If there exist multiple equlibria we seek to characterize whether the answers of the above questions change in different equilibrium strategies.
 
\subsubsection{Results} We fully characterize  the equilibrium (may be multiple) pricing strategies in the push, pull and hybrid model. 

  Our analysis reveals that in the {\em push} model--
\begin{itemize}
\item There is a unique SPNE when the advertisement revenue per user is below a certain threshold (say, $T$), but there are infinitely many SPNEs when it  exceeds $T$ (Theorem~\ref{thm:push}) .
\item The price charged by the IoTSP (payoff of the IoTSP, respectively) decreases (increases,resp.) as the advertisement revenue per user increases when it is below $T$ (Theorem~\ref{thm:push-iot}, Corollary~\ref{cor:push_lowad}, resp.).  At and above $T$, the IoTSP can sustain a service free of cost   (Theorem~\ref{thm:push-iot}). The IoTSP\rq{}s payoff becomes constant when the advertisement revenue per user exceeds $T$ (Corollary~\ref{cor:push_highad}); thus, the IoTSP only retains a fixed share of the advertisement revenue and the rest is shared between the WSP and CSP in this regime.
\item The price (payoff, resp.) of the WSP (or CSP) increases with the advertisement revenue per user when it is below $T$ (Theorem~\ref{thm:push}, Corollary~\ref{cor:push_lowad}, respv. ). At and above $T$, the price (payoff, respv.) will depend on the selection of  equilibrium since there are multiple SPNEs, but the sum of the payoffs of the WSP and CSP is always unique and  increases linearly with the advertisement revenue (Corollary~\ref{cor:push_highad}).
\item Since the demand increases monotonically with the decrease in price quoted by the IoTSP,  the demand increases when the advertisement revenue per user is below $T$ (Corollary~\ref{cor:push_lowad}). At and above $T$, the demand remains constant at the {\em maximum value} (Corollary~\ref{cor:push_highad}). 
\end{itemize}

  Our analysis shows that in the {\em pull} model
\begin{itemize}
\item Unlike the push model, there is a unique NE in the pull model for all sets of parameters (Theorem~\ref{thm:pull}).
\item The payoff of the IoTSP increases monotonically with advertisement revenue per user for all sets of parameters (Corollaries ~\ref{cor:pull_lowad} and~\ref{cor:pull_highad}). Payoffs of the WSP and CSP increase with advertisement revenues per user when it  is below a certain threshold ($T_1$)  (Corollary~\ref{cor:pull_lowad}) and remain constant when it exceeds $T_1$ (Corollary~\ref{cor:pull_highad}). Thus, unlike the push model, the IoTSP grabs all the advertisement revenue when it  becomes very high.
\item Total payment that an end-user has to make decreases (remains constant, resp.) when the advertisement revenue per user is below $T_1$ (at and above $T_1$, resp.) (Theorem~\ref{thm:pull}).
\item The behavior of the demand is similar to the push model (Corollaries~\ref{cor:pull_lowad}$\&$\ref{cor:pull_highad}). Only difference is that when the advertisement revenue per user exceeds $T_1$ then it remains constant at a value {\em lower} than the maximum value (Corollary~\ref{cor:pull_highad}). 
\end{itemize}

Our analysis reveals that in the {\em hybrid} interaction model
\begin{itemize}
\item There  exists a unique SPNE pricing strategy for providers in the hybrid model for all sets of parameters (Theorem~\ref{thm:hybrid}). 
\item Payoffs of all providers increase as the advertisement revenue per user increases when it is below a certain threshold ($T_2$) (Corollary~\ref{cor:hybrid_lowad});  payoffs of the IoTSP and WSP become constant when it exceeds $T_2$ (Corollary~\ref{cor:hybrid_highad}). Thus, the IoTSP  only retains a fixed share from the advertisement revenue and the rest is captured by the CSP in the high advertisement revenue per user regime. 
\item Total payment that an end-user has to make decreases (remains constant, resp.) when the advertisement revenue per user is below $T_2$ (at and above $T_2$, resp.) (Theorem~\ref{thm:hybrid}).
\item The behavior of demand is similar to the push and pull model (Corollaries~\ref{cor:hybrid_lowad} $\&$ \ref{cor:hybrid_highad}). Only difference is that when the advertisement revenue per user exceeds $T_2$ then it remains constant at a value {\em lower} than the maximum value, but it is greater than that of the pull model (Corollary~\ref{cor:pull_highad}).
\end{itemize}

We show that  the IoTSP selects a price which maximizes the advertisement revenue that the IoTSP will get for each end-user in all the models (Section~\ref{game:IoTSP and advertisers}). 

We compare the payoffs and the demand associated with these models (Section~\ref{sec:comparison_models})--
\begin{itemize}
\item  Demand and thus, the reach of the IoT technology, is the highest in the pull (push, resp.) model for low (high, resp.) advertisement revenue per user regime (Corollary~\ref{cor:demand_compare}).  Demand in the hybrid model is the same (lower, resp.) as in the push model in the low (higher, resp.) advertisement revenue per user regime.
\item The IoTSP\rq{}s payoff is the highest in the pull model (Corollary~\ref{cor:iot_compare}) for all sets of parameters. IoTSP\rq{}s payoff in the hybrid model is the same (lower, resp.) as in the push model in the low (high, resp.) advertisement revenue per user regime. 
\item The payoff of the WSP is the highest (lowest, resp.) in the pull model for  low (high, resp.) advertisement revenue per user regime (Corollary~\ref{cor:wsp_compare}). The WSP\rq{}s payoff in the push model is always higher compared to the hybrid model. Thus, the WSP's payoff is the highest in the push model in the high advertisement revenue per user regime.  
\item The payoff of the CSP is the highest in the hybrid model  in the low advertisement revenue  regime (Corollary~\ref{cor:csp_compare}).  The payoff of the CSP is higher (lower, resp.) in the push model compared to the pull model in the high (low, resp.) advertisement revenue regime. In the high advertisement revenue per user regime, the CSP\rq{}s payoff is not unique in the push model because the SPNE is not unique.   We show that the CSP\rq{}s payoff is higher (lower, resp.)  in the hybrid model compared to the worst (best, resp.) possible payoff that the CSP attains in the push model  (Corollary~\ref{cor:csp_compare}).
\item Each participating advertiser will prefer the model which incurs the highest demand (Section~\ref{sec:advertiser_compare}). 
\end{itemize}

\subsection{Related Literature}
To the best of our knowledge this is the first investigation of the economic aspect of interactions among providers of different kinds in an IoT setting.  However, the pricing strategies in a system consisting of only one kind of provider has been extensively studied e.g. ISP \cite{mackie}, WSP \cite{lehr,archived} or CSP\cite{ardanga,pal}.  But different modes of interactions like push, pull and hybrid naturally arise when there are multiple kinds of providers as in the IoT. Payoff functions of different kinds of providers are also different; for example advertisers directly pay only the IoTSP, while the WSP and CSP only get indirect shares of the advertisement revenue either from end-users or the IoTSP.    

Some recent work on the net-neutrality considered the interaction between the ISP and Content provider (CP) \cite{altman_push,altman_pull}. \cite{altman_push} considers a setting similar to the push model where the ISP only charges the CP and  the  CP charges the end-users. \cite{altman_pull} considers that the ISP and CP both charge the end-users (similar to the pull model in our setting). But the pricing game in the IoT context requires a different problem formulation. For example \cite{altman_push} needs a {\em purely} sequential game where first the CP selects its price and then the ISP selects its price; but the push model which is the closest to \cite{altman_push} needs a combination of a sequential and a parallel game; in the first stage of this interaction, the WSP and CSP select their prices in parallel and then the IoTSP selects its price.  The results are also substantially different.  For example, \cite{altman_push} shows that there always exists a unique SPNE. But we find  that there may be infinitely many SPNEs in the push model. We also characterize the interaction between the IoTSP and advertisers and investigate the optimal price that the IoTSP should quote to the advertisers. This enables us to characterize the demand of end-users and the payoffs of the providers in all the possible advertisement revenue regime which have not been studied in \cite{altman_push} and \cite{altman_pull}. Moreover, the hybrid interaction model can only arise when there are more than two different kinds of providers in the system as in the IoT. Hence, the hybrid interaction model has not been studied in the above papers.     We also analytically provide a comparison of the payoffs of providers and demand in different models.   

In Economics, interactions among providers of different kinds have been studied in the supply chain management where different providers select prices for different complementary products \cite{wang, yin}. \cite{wang} studied the model where the providers select prices simultaneously to the end-users (similar to the pull model in our setting).  \cite{yin} studied a model similar to the push model where  in the first stage providers select prices simultaneously (similar to the WSP and CSP) and in the second stage an assembler selects price (similar to the IoTSP) to the end-users for the assembled product. The main differences with the above mentioned papers are twofold.

First, both of the above papers did not consider the impact of the advertisement revenue. The consideration of the above leads to a different characterization of the payoff function of the IoTSP. The IoTSP\rq{}s payoff now depends on the demand of the end-users, the price quoted by the IoTSP and the advertisement revenue for each user whereas in \cite{wang} and \cite{yin} the payoff of each provider only depends on the demand of the end-users and the price quoted by the provider. Since the payoff function is different in our setting,  the analysis of the pricing interaction among the providers is also different. The analysis of the game in the high advertisement revenue regime is also significantly different and challenging compared to the low advertisement revenue regime. For example, when the advertisement revenue per user becomes sufficiently high, then the price quoted by  the IoTSP becomes $0$ in the push model, and the payoff function of the providers is no longer a strictly concave function in the decision variables. Hence, the first order condition is no longer sufficient for the existence of the Nash Equilibrium in the high advertisement revenue regime. Thus, the result is different in the high advertisement revenue regime compared to the lower advertisement revenue regime. For example, we show that there are infinitely many SPNEs whereas there is a unique SPNE in the lower advertisement revenue regime. \cite{yin} also concluded that the the push interaction is not socially optimal since the price quoted to the end-users never becomes $0$. However, our analysis shows that in the high advertisement revenue regime, the payment that each end-user incurs becomes $0$. 

Second, \cite{wang} and \cite{yin} did not consider the hybrid interaction model among the providers which we consider. In the hybrid model, we need to introduce a different economic framework compared to the push and the pull model. In the hybrid interaction model, in the first stage the CSP selects its price to the IoTSP and then the IoTSP and the WSP select their prices simultaneously to the end-users. 

Additionally, \cite{wang} considered a model where the demand is always non zero and the demand becomes infinite when the price becomes $0$. In practice, the demand is always finite and the demand will be zero when the price is sufficiently high which we model in our setting. The consideration of the above significantly altered the characterization of the pricing strategy. 

{\em Proofs have been relegated to the Appendix. Due to the space constraint we relegate some algebraic manipulations in the proofs to our archived report \cite{isit_15}.}
\vspace{-0.1cm}
\section{Players, Decision Variables and Parameters}\label{sec:parameters}
We investigate a monopoly where there is one provider of each kind i.e. there is one IoTSP, WSP and CSP. 

The IoTSP {\em decides} the price $b\geq 0$ for per unit of advertisement volume and in response, each advertiser {\em decides} to advertise or not depending on the advertiser\rq{}s expected revenue that it will get from each end-user. The total advertisement volume is $a_1$ for each end-user.   The IoTSP receives $ba_1$ amount of {\em advertisement revenue per user}. In general, $a_1$ is a decreasing function of $b$. 

  Advertisers {\em decide} the additional cloud resources $a_2$ required for each end-user for providing customized advertisement. $a_2$ is a function of $a_1$. We discuss the interaction between IoTSP and advertisers in detail in Section~\ref{game:IoTSP and advertisers}.

   The IoTSP {\em decides} a price $p_I\geq 0$ to end-users for an IoT application in the push, pull and hybrid model\footnote{There may be an additional fixed installation cost which we do not consider.}. The price may be a periodic subscription fee\footnote{AT$\&$T has already announced monthly subscription scheme for its smart home secure appliances\cite{att}.} or may be the price per application\footnote{One such example is that Amazon charges a price only when a book is downloaded via Kindle. End-users do not need to pay a periodic subscription fee. }. The WSP {\em decides} a charge of $p_w\geq 0$  per data (in bytes). Note that in the push model (the pull and hybrid models, resp.) the charge is levied on the IoTSP (end-users, resp.). We consider a usage based pricing scheme, but our result easily generalizes to other pricing schemes. The CSP {\em decides} a charge of $p_c\geq 0$ for each unit of resource. Note that in the push and hybrid models (pull model, resp.) the charge is levied on the IoTSP (end-users, resp.). Similar type of pricing is currently employed by Amazon EC2\cite{amazon}.

We denote-- i) the  average data flow (in bytes per second) required for each end-user for IoT application as $\alpha$; ii) cloud capacities required to serve one end-user as $\beta$. Note that $\alpha+a_1$ ($\beta+a_2$, resp.)  is the total data traffic   (cloud capacity required, resp.) for each end-user.

We consider that in the first stage of each interaction model the IoTSP selects $b$ and in response, $a_1$ is selected (Fig.~\ref{fig:push_pull_hybrid}, Section~\ref{game:IoTSP and advertisers}). Since $a_2$ is a function of $a_1$,  $a_2$ is readily obtained once $a_1$ is known. {\em After knowing $b, a_1$ and $a_2$} the IoTSP, WSP and CSP select $p_I, p_w, p_c$ respectively in different interaction models (push, pull and hybrid). In the push model, the CSP and the WSP select their prices $p_w, p_c$ to the IoTSP simultaneously and then, the IoTSP selects its price $p_I$ to the end-users (Fig.~\ref{fig:push_pull_hybrid}). In the pull model, the IoTSP, the WSP and CSP select prices $p_I, p_w, p_c$ simultaneously to the end-users (Fig.~\ref{fig:push_pull_hybrid}). In the hybrid model, the CSP selects its price $p_c$ to the IoTSP and then the IoTSP and the WSP selects prices $p_I$ and $p_w$ respectively to the end-users (Fig.~\ref{fig:push_pull_hybrid}). 

We characterize the equilibrium prices, the payoffs of the providers and the demand of the end-users by considering $b$ and $a_1$ as parameters in the push (Section~\ref{sec:push}), the pull (Section~\ref{sec:pull}) and the hybrid (Section~\ref{sec:hybrid}) model.  Subsequently, we analyze how IoTSP should select $b$  in order to maximize its own payoff in Section~\ref{game:IoTSP and advertisers} in each of the interaction models.

\vspace{-0.1cm}
\section{Push Model}\label{sec:push}

\subsection{System Model}\label{sec:push_model}
In the push model,  the price of the IoTSP {\em directly} impacts the demand of end-users. The prices of the WSP and CSP {\em directly} impact the prices of the IoTSP. Thus, it is reasonable to assume that whenever the WSP and CSP vary their prices, the IoTSP will also change its own price. Thus, we consider a non cooperative game where in the first stage the WSP and CSP decide their prices $p_w, p_c$ they will charge  the IoTSP  (sub-game 1)  and in the last stage  the IoTSP selects the price $p_I$ to the end-users with the knowledge of prices of the WSP and the CSP (sub-game 2) (Fig.~\ref{fig:push_pull_hybrid}). The overall game is a variant of {\em Stackelberg game}\footnote{In a  Stackelrberg kind of game\cite{mwg}, the leader first selects its action, and then the follower selects its action with the complete knowledge of the action chosen by the leader. In the push interaction, however, there are multiple followers, hence it is different than traditional Stackelberg game.}where the CSP and WSP are the leaders and the IoTSP is the follower.

The demand of end-users only depends on the price of IoTSP i.e. $p_I$.
We assume that the demand-response of end-users follows a linear relationship-- 
\begin{align}\label{demand}
D_{push}=\max\{D_{max}-dp_I,0\}
\end{align}
 where $d$ denotes the price sensitivity\footnote{High price sensitivity indicates that demand will decrease (increase, resp.) at a high rate with an increase (decrease, resp.) in the price.} of end-users, $D_{max}$ is the maximum demand. A linear demand response function is prevalent in practice and has been extensively studied in Economics\cite{linear}, and internet \cite{altman_pull}. The consideration of the non-linear demand response function is a work for the future.

 The IoTSP receives  $D_{push}ba_1$ amount from the advertisement, but, it also has to pay $p_w(\alpha+a_1)D_{push}$ amount to the WSP and $p_c(\beta+a_2)D_{push}$ amount to the CSP. Thus, the IoTSP\rq{}s payoff is
\begin{align}
U_{I,push}=p_ID_{push}+bD_{push}a_1-p_wD_{push}(\alpha+a_1)\nonumber\\
-p_cD_{push}(\beta+a_2)\label{utilitypush}\\
\text{WSP\rq{}s payoff }=U_{w,push}=p_wD_{push}(\alpha+a_1)\label{utilitywsppush}\\
\text{CSP\rq{}s payoff }=U_{c,push}=p_cD_{push}(\beta+a_2)\label{utilitycsppush}
\end{align}
  
We seek to characterize the Sub-game perfect Nash equilibrium (SPNE) of the above game.
\begin{defn}\cite{mwg}
A Sub-game perfect Nash equilibrium (SPNE) strategy profile  is an NE \cite{mwg} at every subgame.
\end{defn}
In Section~\ref{sec:push_result} we characterize the equilibrium prices $p_w, p_I, p_c$ and the payoffs of providers in terms of $b, a_1$ and $a_2$, subsequently in Section~\ref{game:IoTSP and advertisers} we will show how the IoTSP should select $b$.
\subsection{Results}\label{sec:push_result}
We summarize the main results of this section--
\begin{itemize}
\item The NE pricing strategies of the WSP and CSP  (i.e. at sub-game 1) are {\em unique} when the advertisement revenue per user, $ba_1\leq 5D_{max}/d$ , however {\em there are infinitely many NE pricing strategies} when $dba_1>5D_{max}$ (Theorem~\ref{thm:push}). The NE pricing strategy of the IoTSP in the sub-game $2$ is always unique (Theorem~\ref{thm:push-iot}). When $dba_1>5D_{max}$, though there are infinitely many SPNEs, the IoTSP\rq{}s payoff as well as the sum of the payoffs of the WSP and CSP are always unique (Corollary~\ref{cor:push_highad}).
\item The price of the IoTSP quoted to the end-users decreases as $ba_1$ increases and becomes $0$ when $dba_1\geq 5D_{max}$\footnote{Note that similar pricing strategy is also prevalent in some legacy technologies. For example,  in the app market, some application providers sell their apps at a free of cost, but it also comes with a lot of advertisement; on the other hand,  some application providers charge the end-users for their products, but these apps do not provide a lot of advertisement.}. Thus, even though the IoTSP has to share its revenue with the CSP and WSP, it can still offer its service at free.  Intuitively, the total advertisement revenue procured by the IoTSP is $ba_1D_{push}$; thus, when $ba_1$ is  high, then the IoTSP can procure high advertisement revenue  by selecting a lower price which in turn increases $D_{push}$, and the revenue. 

The prices of the WSP and CSP quoted to the IoTSP increase with $ba_1$ when $dba_1\leq 5D_{max}$.  Though there are infinitely many equilibria when $dba_1>5D_{max}$,  the sum of payment received by WSP and CSP for each end-user is unique and increases linearly  with $ba_1$.
\item Since the price quoted by the IoTSP decreases (Theorem~\ref{thm:push-iot}) with $ba_1$ when $dba_1\leq 5D_{max}$, the demand increases in this regime (Corollary~\ref{cor:push_lowad}).   The demand becomes equal to the maximum possible value since the price charged by the IoTSP to the end-users becomes $0$ when $dba_1\geq 5D_{max}$(Corollary \ref{cor:push_highad}).

\item The payoffs of the providers depend on the demand and the prices. We now summarize the payoffs at the equilibrium. We show that the payoffs of the providers increase  with $ba_1$ (Fig.\ref{fig:push_pay}) when $ba_1\leq 5D_{max}$ (Corollary~\ref{cor:push_lowad}).  Thus, if the IoTSP can procure high $ba_1$, then, it not only increases the payoff of the IoTSP, but it also increases the payoffs of the WSP and CSP though the advertisers only pay the IoTSP. The WSP and CSP attain  twice the payoff of the IoTSP (Corollary~\ref{cor:push_lowad}). Intuitively, the prices of the WSP and CSP do not {\em directly} affect the demand of end-users, thus, they can gain more by selecting higher prices even though the IoTSP has direct source of revenue from advertisers. 

 The payoff of the IoTSP is independent of $ba_1$ in this regime (from (\ref{utilityiotsphighad}), Fig.\ref{fig:push_pay}). Thus, the IoTSP retains only a constant amount of revenue from the advertisement. The rest of the advertisement revenue is shared between the CSP and the WSP ( by \ref{utility_sum_wsp_csp})).  Thus, the WSP and the CSP receive higher shares of the advertisement revenue even though the advertisers {\em only} pay the IoTSP.

\end{itemize}
We now describe the results in detail.
\subsubsection{Equilibrium Price}

\textbf{NE Pricing $p_I^{*}$ at sub-game 2}: 
\begin{Lemma}\label{lm:push}
The unique NE pricing strategy $p_I^{*}$ in sub-game 2 is
\begin{align}\label{iotpricepush}
p_I^{*}=\max\left\{\dfrac{D_{max}}{2d}-\dfrac{ba_1}{2}+\dfrac{p_w(\alpha+a_1)}{2}+\dfrac{p_c(\beta+a_2)}{2},0\right\}.
\end{align}
\end{Lemma}
\textbf{NE Pricing strategy  $(p_w^{*},p_c^{*})$ at sub-game 1}:
\begin{Theorem}\label{thm:push}
When $dba_1<5D_{max}$, then the unique NE $(p_w^{*},p_c^{*})$ in the sub-game is
\begin{align}
p_w^{*}=\dfrac{D_{max}}{3d(\alpha+a_1)}+\dfrac{ba_1}{3(\alpha+a_1)}\label{eq:wsppricelowad}\\
p_c^{*}=\dfrac{D_{max}}{3d(\beta+a_2)}+\dfrac{ba_1}{3(\beta+a_2)}\label{eq:csppricelowad}
\vspace{-0.1cm}
\end{align}

When $dba_1\geq 5D_{max}$, then any $(p_w^{*},p_c^{*})$ which satisfies the following conditions constitutes an NE in this sub-game
\begin{align}
p_w^{*}(\alpha+a_1)\in [2D_{max}/d,ba_1-3D_{max}/d]\label{eq:wsppricehighad}\\
p_c^{*}(\beta+a_2)\in [2D_{max}/d,ba_1-3D_{max}/d]\label{eq:csppricehighad}\\
\text{such that } p_c(\beta+a_2)+p_w(\alpha+a_1)=ba_1-D_{max}/d\label{eq:highadcondition}
\end{align}
\vspace{-0.2cm}
 \end{Theorem} 
 Note that $ba_1-3D_{max}/d\geq 2D_{max}/d$ when $dba_1\geq 5D_{max}$. Thus, the  interval from which the WSP and  CSP select their prices given in (\ref{eq:wsppricehighad}) and (\ref{eq:csppricehighad}) respectively is non-empty.  
 
Replacing the values of $(p_w^{*},p_c^{*})$ in (\ref{iotpricepush}): 
\begin{Theorem}\label{thm:push-iot}
At SPNE,
\begin{eqnarray}
p_I^{*}=\max\left\{\dfrac{5D_{max}}{6d}-\dfrac{ba_1}{6},0\right\}\label{eq:iotprice}
\vspace{-0.1cm}
\end{eqnarray}
\end{Theorem}
The above theorem entails that though $(p_w^{*},p_c^{*})$ are not always unique, $p_I^{*}$ is always unique.
\subsubsection{Payoffs of providers and Demand of end-users}
-\\
a) $dba_1<5D_{max}$:
\begin{corollary}\label{cor:push_lowad}
When $dba_1< 5D_{max}$, then at equilibrium
\begin{eqnarray}
D_{push}
& =\dfrac{D_{max}}{6}+d\dfrac{ba_1}{6}\nonumber\\
U_{I,push}& =d\left(\dfrac{D_{max}}{6d}+\dfrac{ba_1}{6}\right)^2\nonumber\\
U_{w,push}=U_{c,push}
& =2d\left(\dfrac{D_{max}}{6d}+\dfrac{ba_1}{6}\right)^2\nonumber
\end{eqnarray}
\end{corollary}

b) $dba_1\geq 5D_{max}$: Now we evaluate the demand and payoffs at equilibrium when $dba_1\geq 5D_{max}$. Since the SPNE is not unique in this case we also characterize the worst possible payoffs for the WSP ($U_{w,worst,push}$) and the CSP ($U_{c,worst,push}$) and the best possible payoffs for the WSP ($U_{w,best,push}$) and the CSP ($U_{c,best,push}$).  
\begin{corollary}\label{cor:push_highad}
When $dba_1\geq 5D_{max}$, then
\begin{eqnarray}\label{eq:demandhighad}
& D_{push}= D_{max}\nonumber\\
& U_{I,push}
=D_{max}^2/d, \quad {\text{at each SPNE}}\label{utilityiotsphighad}\\
& U_{w,worst,push}=U_{c,worst,push}= \dfrac{2D_{max}^2}{d} \quad \nonumber\\
& U_{c,best,push}=U_{w,best,push}= (ba_1-3D_{max}/d)D_{max}\nonumber\\
& U_{w,push}+U_{c,push}=ba_1D_{max}-D_{max}^2/d\label{utility_sum_wsp_csp}
\end{eqnarray}
\end{corollary}


Fig.~\ref{fig:push_pay} shows the variation of the IoTSP and the sum of the payoffs of the WSP and CSP with $ba_1$ in an example setting. 

\section{Pull Model}\label{sec:pull}
\subsection{System Model}\label{sec:pull_model}
 The IoTSP, WSP and CSP all directly charge prices to the end-users in the pull model (Fig.~\ref{fig:push_pull_hybrid}). An end-user needs to pay $p_w(\alpha+a_1)$ and $p_c(\beta+a_2)$ amount to the WSP and CSP respectively. Thus, the demand is
\begin{align}\label{demandpull}
D_{pull}=\max\{D_{max}-d(p_I+p_w(\alpha+a_1)+p_c(\beta+a_2)),0\}
\end{align}
 We consider the same price sensitivity parameter $d$ for the prices of different providers\footnote{The generalization of our model to account for different price sensitivity parameters for different providers is straightforward.}. 
 
Since the IoTSP does not pay either the WSP or CSP, the payoff functions are
\begin{align}
U_{I,pull}=p_I*D_{pull}+ba_1*D_{pull}\label{utilitypull}\\
\text{WSP\rq{}s payoff }=U_{w,pull}=p_w(\alpha+a_1)D_{pull}\label{utilitypullwsp}\\
\text{CSP\rq{}s payoff }= U_{c,pull}=p_c(\beta+a_2)D_{pull}\label{utilitypullcsp}
\end{align}
  Since each provider independently quotes its price to the end-users, we consider a non cooperative game where each provider {\em simultaneously} selects its price.  We characterize the NE strategy profile in terms of $b, a_1$ and $a_2$. 
\begin{figure*}
\begin{minipage}{.32\linewidth}
\begin{center}
\includegraphics[width=60mm,height=30mm]{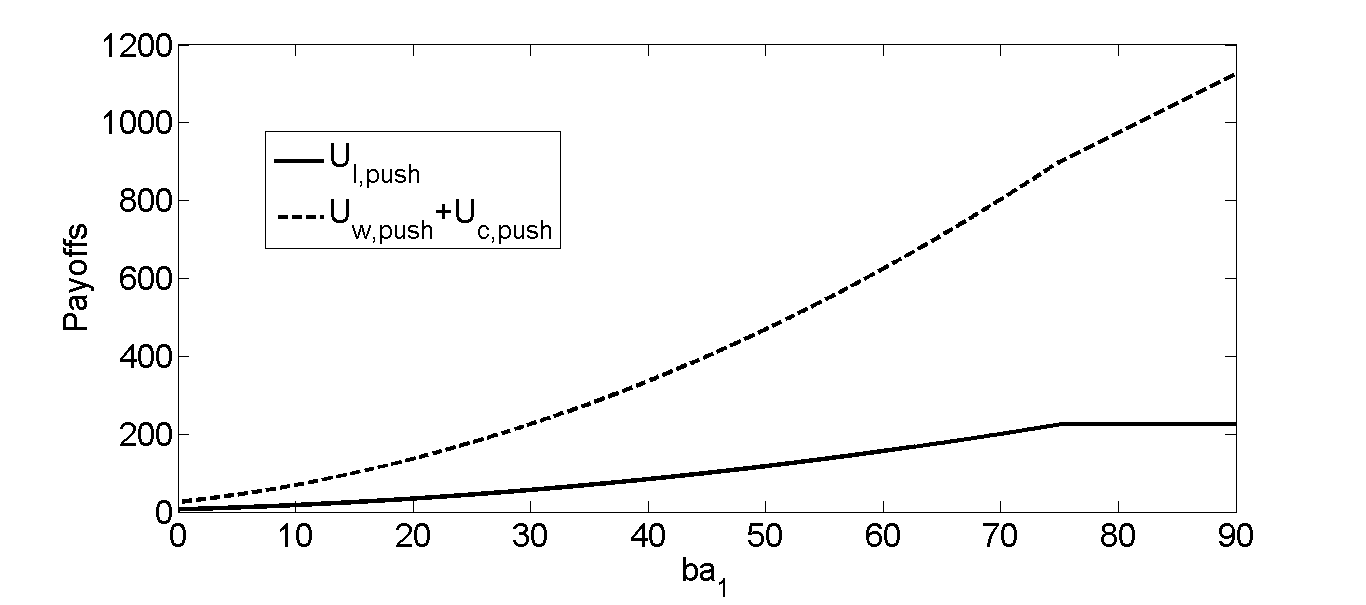}
\vspace{-.5cm}
\caption{We consider: $d=1, D_{max}=15$. $U_{I,push}$ becomes static when $dba_1\geq 5D_{max}=75$, but $U_{c,push}+U_{w,push}$ linearly increases when $ba_1\geq 75$.}
\label{fig:push_pay}
\end{center}
\end{minipage}\hfill
\begin{minipage}{.32\linewidth}
\begin{center}
\includegraphics[width=60mm,height=30mm]{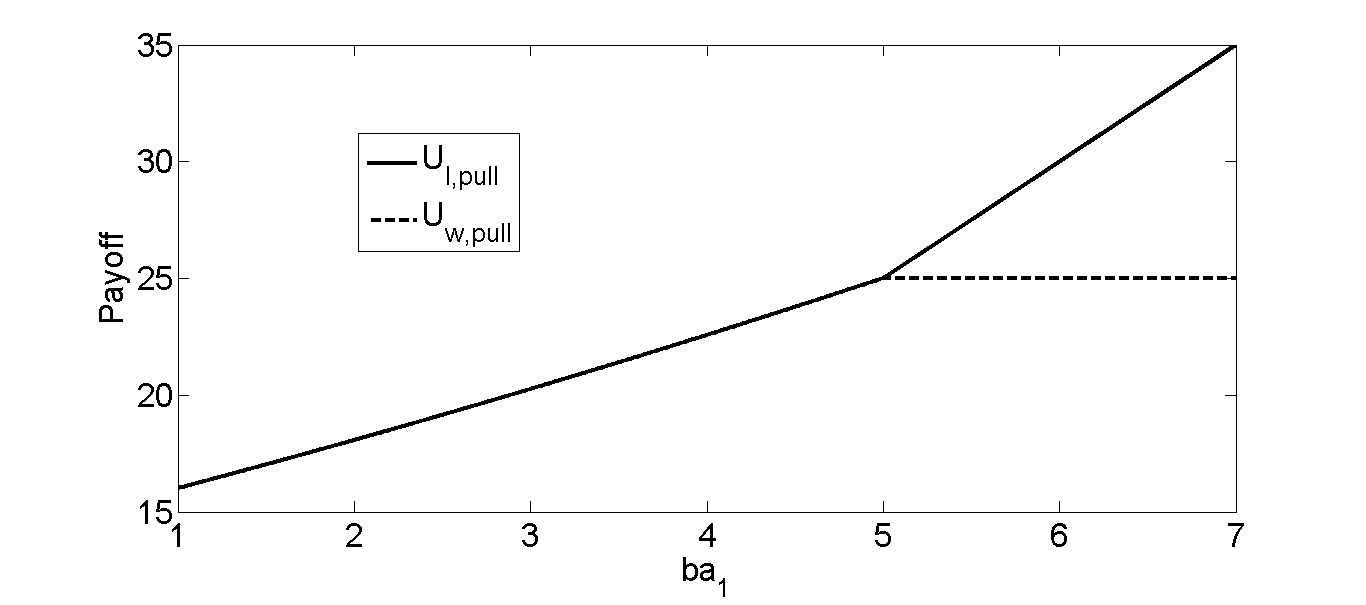}
\vspace{-.5cm}
\caption{We consider the same example setting as in fig.~\ref{fig:push_pay}. $U_{I,pull}$ and $U_{w,pull}$ are the same for $ba_1\leq 5$, but when $ba_1>5$, $U_{I,pull}>U_{w,pull}$. From Corollaries~\ref{cor:pull_lowad} and \ref{cor:pull_highad} $U_{w,pull}=U_{c,pull}$.}
\label{fig:pull_pay}
\end{center}
\end{minipage}\hfill
\begin{minipage}{.32\linewidth}
\begin{center}
\includegraphics[width=60mm,height=30mm]{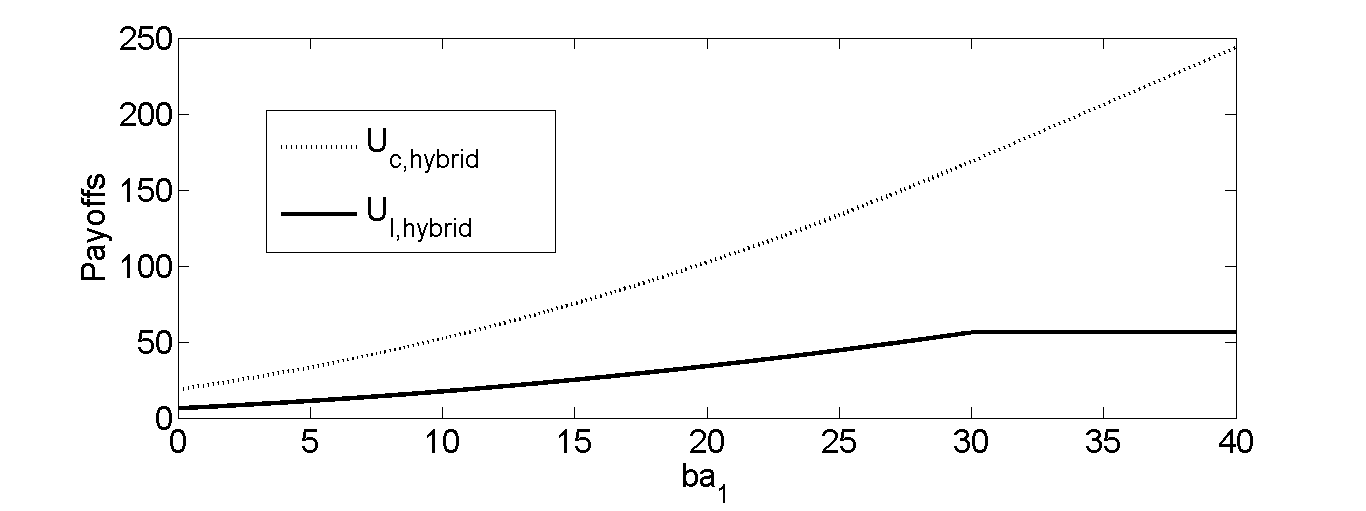}
\vspace{-.5cm}
\caption{We consider the same example setting as in fig.~\ref{fig:push_pay}. $U_{I,hybrid}$ remains constant when $ba_1\geq 2D_{max}/d=30$, but $U_{c,hybrid}$ always increases with $ba_1$. From Corollaries~\ref{cor:hybrid_lowad} and \ref{cor:hybrid_highad} note that $U_{I,hybrid}=U_{w,hybrid}$.}
\label{fig:hybrid_pay}
\end{center}
\end{minipage}
\vspace*{-.6cm}
\end{figure*}
\subsection{Results}\label{sec:pull_result}
We, first, summarize the main attributes of the NE strategy and subsequently, we describe the results in detail.
\begin{itemize}
\item Unlike in the push model, the NE pricing strategy is always unique  in the pull model (Theorem~\ref{thm:pull}). 
\item Similar to the push model,  the price quoted to the end-users by the IoTSP decreases with $ba_1$ (recall that $ba_1$ is the advertisement revenue per user) and the prices of the WSP and CSP quoted to the end-users increase with $ba_1$  when $ba_1$ is below a threshold. However, the threshold is $D_{max}/(3d)$ in the pull model whereas the threshold is $5D_{max}/d$ in the push model. The reduction in the price by the IoTSP helps the WSP and CSP to increase their prices without affecting the demand much. The price selected by the IoTSP becomes $0$ when $dba_1\geq D_{max}/3$.  The prices of the WSP and CSP also become independent of $ba_1$. Once $p_I$ becomes $0$, the IoTSP can not decrease its price further, thus, the WSP and CSP can not increase their prices as it will decrease the demand which in turn reduces the payoffs. 
\item The demand $D_{pull}$ increases with $ba_1$ despite that end-users pay the WSP for the additional advertisement traffic $a_1$ (Corollary~\ref{cor:pull_lowad}). Note from Theorem~\ref{thm:pull} that the increase in $ba_1$ enables the IoTSP to decrease its price which enhances the demand.  Corollary~\ref{cor:pull_highad} entails that the demand becomes independent of $ba_1$ when $dba_1>D_{max}/3$. This is because the total payment that an end-user incurs becomes independent of $ba_1$ when $dba_1>D_{max}/3$ (by Theorem~\ref{thm:pull}).  Each end-user has to pay the WSP and CSP under all cases unlike in the push model, thus, the demand never reaches the maximum possible value, $D_{max}$.

\item The payoffs are the same for each provider when $dba_1\leq D_{max}/3$ unlike in the push model (Corollary~\ref{cor:pull_lowad}, Fig.\ref{fig:pull_pay}).

Since the payment made by an end-user becomes independent of $ba_1$, the payoffs of the WSP and CSP become independent of $ba_1$ when $dba_1>D_{max}/3$.  The IoTSP\rq{}s payoff increases linearly with $ba_1$ when $dba_1>D_{max}/3$ (Corollary~\ref{cor:pull_highad}).  Unlike in the push model, the IoTSP retains all the revenue generated from advertisement when the advertisement revenue per user is high (Corollary~\ref{cor:pull_highad}, Fig.\ref{fig:pull_pay}). Hence in this regime, the WSP and CSP do not get any share of the advertisement revenue  unlike in the push model. Intuitively, in the push model, the WSP and CSP quote their prices to the IoTSP; thus, the CSP and the WSP can  increase their prices with the increase in $ba_1$ without {\em directly} decreasing the demand, hence the payoffs of the CSP and WSP always increase with $ba_1$. However, in the pull model, the WSP and the CSP quote their prices to the end-users, thus, once the price quoted by the IoTSP becomes $0$, the WSP and CSP can not increase their prices without decreasing the demand since the IoTSP can not decrease its price below $0$.  
\end{itemize}
\vspace{-0.1cm}
\subsubsection{Equilibrium prices $(p_I^{*},p_w^{*},p_c^{*})$}
\begin{Theorem}\label{thm:pull}
The NE strategy profile is unique--
\begin{eqnarray}
p^{*}_I& =\begin{cases}
\dfrac{D_{max}}{4d}-\dfrac{3ba_1}{4}\quad\mbox{if } dba_1\leq D_{max}/3\nonumber\\
0 \quad \mbox{if } dba_1>D_{max}/3
\end{cases}\\
p^{*}_w& =\begin{cases}\dfrac{D_{max}}{4d(\alpha+a_1)}+\dfrac{ba_1}{4(\alpha+a_1)}\quad \mbox{if } dba_1\leq D_{max}/3\nonumber\\
 \dfrac{D_{max}}{3d(\alpha+a_1)}\quad \mbox{if } dba_1>D_{max}/3
\end{cases}
\label{wsppricelowadpull}\\
p^{*}_c& =\begin{cases} \dfrac{D_{max}}{4d(\beta+a_2)}+\dfrac{ba_1}{4(\beta+a_2)}\quad \mbox{if } dba_1\leq D_{max}/3\nonumber\\
\dfrac{D_{max}}{3d(\beta+a_2)}\quad \mbox{if } dba_1>D_{max}/3\end{cases}\label{csppricelowadpull}
\end{eqnarray}
\end{Theorem}
Note that the NE strategy profile is always unique unlike in the push model. 
\subsubsection{Payoffs of providers and Demand of end-users}
\paragraph{$dba_1\leq D_{max}/3$}
\begin{corollary}\label{cor:pull_lowad}
When $dba_1\leq D_{max}/3$, then at equilibrium
\begin{eqnarray}\label{demandlowad_pull}
D_{pull}& 
=\dfrac{D_{max}}{4}+d\dfrac{ba_1}{4}\nonumber\\
U_{I,pull}=U_{w,pull}=U_{c,pull}& =d\left(\dfrac{D_{max}}{4d}+\dfrac{ba_1}{4}\right)^2\nonumber
\end{eqnarray}
\end{corollary}


\paragraph{$dba_1>D_{max}/3$}
\begin{corollary}\label{cor:pull_highad}
When $dba_1>D_{max}/3$, then at equilibrium
\begin{eqnarray}
D_{pull}=&\dfrac{D_{max}}{3}\quad \nonumber\\
U_{I,pull}=& ba_1\dfrac{D_{max}}{3}\label{eq:iotpullhighadpay}\\
U_{w,pull}=U_{c,pull}=& D_{max}^2/(9d)\nonumber
\end{eqnarray}
\end{corollary}
 The payoffs of the IoTSP, WSP and CSP have been depicted in Fig.~\ref{fig:pull_pay}.

%

%
\section{Hybrid Model}\label{sec:hybrid}
\subsection{System Model}
The IoTSP and WSP independently quote their prices to the end-users and the CSP only quotes its price to the IoTSP (Fig.~\ref{fig:push_pull_hybrid}) in this model. Thus, the prices of the IoTSP and WSP directly impact the end-users and the CSP's price directly impacts the IoTSP. Thus, we consider a sequential non cooperative game where in the first stage (Sub-game 1) the CSP quotes its price to the IoTSP, then in the second stage (Sub-game 2) the IoTSP and WSP select their prices  in {\em parallel} with the knowledge of the price of the CSP.  The overall game is a variant of Stackelberg game where the CSP is the leader and the IoTSP, and WSP are the followers.

The demand of the end-users is
\begin{align}\label{demandhybrid}
D_{hybrid}=\max\{D_{max}-d(p_I+p_w(\alpha+a_1)),0\}
\end{align}

In the hybrid model, since the IoTSP has to pay the CSP to procure resources, the IoTSP\rq{}s payoff is 
\begin{align}
U_{I,hybrid}=D_{hybrid}(p_I-p_c(\beta+a_2)+ba_1)\label{utilityhybrid}\\
\text{WSP\rq{}s payoff}=U_{w,hybrid}=p_wD_{hybrid}(\alpha+a_1)\label{utilityhybridwsp}\\
\text{CSP\rq{}s payoff}=U_{c,hybrid}=p_cD_{hybrid}(\beta+a_2)\label{utilityhybridcsp}
\end{align}

In Section~\ref{sec:hybrid_result} we find the SPNE pricing strategy profile and payoffs of providers in terms of $b$, $a_1$ and $a_2$. In Section~\ref{game:IoTSP and advertisers} we discuss how the IoTSP should select $b$.
\subsection{Results}\label{sec:hybrid_result}
We first summarize the main attributes of the equilibrium strategy focussing on the similarities and dissimilarities comapared to the push and pull models.
\begin{itemize}
\item We show that  unlike in the push model (but, similar to the pull model), the SPNE pricing strategy is unique in the hybrid model (Theorems~\ref{thm:csp_hybrid} and \ref{thm:hybrid}).  
\item   Similar to the push and pull models, $p_I$ quoted to the end-users decreases with the advertisement revenue per user, $ba_1$  and $p_w, p_c$ increase with $ba_1$ when $dba_1$ is below a threshold. However, we show that the threshold is $2D_{max}$ which is in between the push ($5D_{max}$) and the pull model ($D_{max}/3$).  When $dba_1\geq 2D_{max}$, $p_I$ becomes $0$.  $p_w$ the price quoted to the end-users by WSP becomes independent of $ba_1$ when $dba_1\geq 2D_{max}$, thus, the payment that an end-user incurs also becomes independent of $ba_1$ when $dba_1\geq 2D_{max}$. However, the price quoted by the CSP to the IoTSP i.e. $p_c$ increases linearly with $ba_1$ even when $dba_1\geq 2D_{max}$.
\item  Similar to the push and pull models, the demand $D_{hybrid}$ is strictly increasing in $ba_1$ when $ba_1$ is below the threshold ($2D_{max}/d$) (Corollary~\ref{cor:hybrid_lowad}) and $D_{hybrid}$ becomes constant when $ba_1$ is large enough (Corollary~\ref{cor:hybrid_highad}). 

However, there are some differences compared to the push and pull models. We show that the demand  is higher (lower, resp.) than the maximum  demand achieved in the pull (push, resp.) model. Intuitively,  since end-users always have to pay the WSP, unlike in the push model the demand never reaches the maximum value; but since end-users do not have to pay the CSP,  unlike in the pull model, the demand remains constant at $D_{max}/2$ which is higher compared to maximum demand achieved in the pull model.

\item The payoffs of the IoTSP and WSP are equal, but the CSP\rq{}s payoff is strictly higher than that of the IoTSP and WSP (Corollary~\ref{cor:hybrid_lowad}). The price of the CSP does not affect {\em directly} the demand of the end-users. Thus, the CSP acquires more payoff compared to the IoTSP and WSP by selecting a higher price but without reducing the demand. Note that in contrast, the payoffs of the IoTSP, WSP and CSP are equal in the pull model. On the other hand, the payoffs of the WSP and CSP are equal and higher compared to the IoTSP in the push model. 

   The payoffs of the IoTSP and the WSP are independent of $ba_1$ when $ba_1\geq 2D_{max}/d$ (Corollary~\ref{cor:hybrid_highad}, Fig.\ref{fig:hybrid_pay}).  The CSP procures the rest of  the revenue from advertising sources when $dba_1\geq 2D_{max}$ (Corollary~\ref{cor:hybrid_highad}, Fig.~\ref{fig:hybrid_pay})  though the advertisers only pay the IoTSP.  Corollary~\ref{cor:hybrid_highad} also shows that the CSP\rq{}s payoff is strictly higher compared to the payoffs of the IoTSP and WSP. In the hybrid model, similar to the push model, the CSP charges the IoTSP. Thus, in the hybrid model, the CSP can increase its price without {\em directly} decreasing the demand, hence, similar to the push model (unlike in the pull model) the CSP procures a greater share of revenue from the advertisement. On the other hand, similar to the pull model (and unlike in the push model), the WSP quotes its price to the end-users in the hybrid model. Thus, once the price quoted by the IoTSP becomes $0$, the WSP can not increase its price without decreasing the demand since the IoTSP can not decrease its price below $0$. Hence, similar to the pull model (unlike in the push model), the WSP\rq{}s payoff becomes independent of $ba_1$ in the high advertisement revenue regime.
\end{itemize}
We now describe the results in detail.
\subsubsection{Equilibrium Prices}

\textbf{NE strategy $(p_I^{*},p_w^{*})$ at the sub-game 2}: 
\begin{Lemma}\label{lem:pwpi}
Under the unique NE at sub-game 2, 

If $p_c(\beta+a_2)>ba_1-\dfrac{D_{max}}{2d}$, then 
\begin{eqnarray}
p^{*}_I=\min\left\{
\dfrac{D_{max}}{3d}-\dfrac{2ba_1}{3}+\dfrac{2p_c(\beta+a_2)}{3},\dfrac{D_{max}}{d}\right\}\label{eq:iotsphybrid1}\\
p^{*}_w=\max\left\{\dfrac{D_{max}}{3d(\alpha+a_1)}+\dfrac{ba_1}{3(\alpha+a_1)}-\dfrac{p_c(\beta+a_2)}{3(\alpha+a_1)},0\right\}\label{eq:wsphybrid1}
\end{eqnarray}
If $ba_1-\dfrac{D_{max}}{2d}\geq p_c(\beta+a_2)$, then 
\begin{eqnarray}
p^{*}_I=& 0, \quad p^{*}_w=& \dfrac{D_{max}}{2d(\alpha+a_1)}\label{eq:iotsphybrid3}
\end{eqnarray}
\end{Lemma}
The upper limit on $p_I^{*}$ appears because the demand becomes $0$ when $p_I$ is higher than $D_{max}/d$.


\textbf{Pricing strategy of the CSP in sub-game 1}:
\begin{Theorem}\label{thm:csp_hybrid}
Under the unique NE in the sub-game 1,

When $dba_1<2D_{max}$,
\begin{align}\label{csppricehybridlowad}
p_c^{*}=\dfrac{ba_1}{2(\beta+a_2)}+\dfrac{D_{max}}{2d(\beta+a_2)}
\end{align}
and when $dba_1\geq 2D_{max}$
\begin{align}\label{csppricehybridhighad}
p_c^{*}=\dfrac{ba_1}{\beta+a_2}-\dfrac{D_{max}}{2d(\beta+a_2)}
\end{align}
\end{Theorem}

%
Thus, from Lemma~\ref{lem:pwpi} and Theorem~\ref{thm:csp_hybrid} we have
\begin{Theorem}\label{thm:hybrid}
Under the unique SPNE
\begin{align}
p^{*}_I& =\begin{cases}\dfrac{2D_{max}}{3d}-\dfrac{ba_1}{3}\mbox{if } dba_1<2D_{max}\\
0, \mbox{if } dba_1\geq 2D_{max}\end{cases}\label{iotpricehybridlowad}\\
p^{*}_w& =\begin{cases}\dfrac{D_{max}}{6d(\alpha+a_1)}+\dfrac{ba_1}{6(\alpha+a_1)}\mbox{if } dba_1<2D_{max}\\
\dfrac{D_{max}}{2d(\alpha+a_1)}\mbox{if } dba_1\geq 2D_{max}\end{cases}\label{wsppricehybridhighad}
\end{align}
\end{Theorem}
Note that here the SPNE is always unique similar to the pull model (unlike in the push model). 

\subsubsection{Payoffs of entities}
\begin{corollary}\label{cor:hybrid_lowad}
When $dba_1<2D_{max}$:
\begin{align}
D_{hybrid}
& =\dfrac{D_{max}}{6}+d\dfrac{ba_1}{6}\nonumber\\
U_{I,hybrid}=U_{w,hybrid}
& =d\left(\dfrac{D_{max}}{6d}+\dfrac{ba_1}{6}\right)^2\nonumber\\
U_{c,hybrid}& =3d\left(\dfrac{D_{max}}{6d}+\dfrac{ba_1}{6}\right)^2\nonumber
\end{align}
\end{corollary}
\begin{corollary}\label{cor:hybrid_highad}
When $dba_1\geq 2D_{max}$:
\begin{align}
D_{hybrid}& =D_{max}/2 \nonumber\\
U_{I,hybrid}=U_{w,hybrid}& =\dfrac{D_{max}^{2}}{4d}\label{utilityiotsphybridhighad1}\\
U_{c,hybrid}& =(ba_1-\dfrac{D_{max}}{2d})D_{max}/2\nonumber
\end{align}
\end{corollary}

Fig.~\ref{fig:hybrid_pay} shows the variation of the payoffs of the IoTSP, WSP and CSP with $ba_1$.
\vspace{-0.1cm}
\section{Interaction between the IoTSP and advertisers}\label{game:IoTSP and advertisers}
Here, we describe the  interaction between the IoTSP and advertisers. We first describe how $a_1$ is obtained. The computing resources $a_2$ required for advertisement is a function of $a_1$.  Hence, $a_2$ is readily obtained once $a_1$ is selected. 

There are $M$ advertising firms who are interested in advertising. An advertising firm may or may not advertise. If an advertising firm decides to advertise, then it advertises a volume of $c$ unit to each end-user.   Without loss of generality, we assume that $c=1$. Hence, the total advertising volume to each end-user, $a_1$ is simply the number of advertising firms who advertise in the IoTSP.  An advertiser also has to pay $b$ amount to the IoTSP. {\em The advertising firm does not pay the WSP or CSP}. The IoTSP first {\em decides} $b$, then, a firm decides to advertise or not. A firm will only advertise if its expected revenue\footnote{The expected revenue in general depends on the probability that an end-user will buy the product and the price of the product.} is greater than $b$. {\em Note that advertising firms may have different valuations}, however they advertise the same amount\footnote{We relax this assumption in the future.} $c$. Hence, the number of firms which will advertise is a decreasing function of $b$. We assume that the number of advertising firms or $a_1$ is given by
\begin{align}\label{eq:advertising}
a_1=\min(A_{max}, G(b))
\end{align}
 where $A_{max}$ is the maximum advertisement allowed which arises because the end-users do not prefer advertisement volume above a threshold, $G(\cdot)$ is a decreasing function.  We also assume that $G(b_{max})=0$. Thus, there is an upper limit on $b$. Since we assume that $a_1$ and $b$ are selected from a closed interval,  $ba_1$ has at least one {\em maximum}.

 The advertising model that we consider is motivated by some of the  existing models in the legacy technologies such as the internet. For example, i) in the current internet model, end-users have to pay the WSP for advertisement volume (e.g. Youtube, Netflix).  The advertisers only pay ($b$) the content providers (e.g. Youtube, Netflix).  ii) In the app market, advertisers pay the app developers the price ($b$) quoted by them, the end-users have to pay the WSP for the advertising volume. In both the models,  the advertisers whose valuations are larger than $b$ participate in the advertising, the rest do not. The advertising volume ($c$) is also often the same for each advertising firm.



Now, we characterize the optimal pricing strategy $b$ of IoTSP for online advertisers.  The IoTSP sets its price $b$ with the knowledge of (\ref{eq:advertising}). 

We first consider the push model. The payoff of the IoTSP increases monotonically with $ba_1$ when $ba_1\leq 5D_{max}/d$ (Corollary~\ref{cor:push_lowad}). However, the payoff of the IoTSP becomes independent of $ba_1$ when $ba_1>5D_{max}/d$ (Corollary~\ref{cor:push_highad}).  Thus, any $ba_1\geq 5D_{max}/d$ will maximize the payoff of the IoTSP. Hence, there can be multiple $b$ which maximize the payoff of the IoTSP. However, if the maximum value of $ba_1$ is less than $5D_{max}$, then the payoff of the IoTSP is maximized {\em only at the maximum value of} $ba_1$. 

 In the pull model, the payoff of IoTSP always monotonically increases with $ba_1$ (Corollaries~\ref{cor:pull_lowad} and \ref{cor:pull_highad}). Hence, only the maximum value of $ba_1$ will maximize the payoff of the IoTSP. If there is a unique maximizer of $ba_1$, then there is a unique $b$ which maximizes the payoff of the IoTSP.  On the other hand, if there are more than one $b$ which maximize the $ba_1$, then all those $b$ will also maximize the payoff of the IoTSP. 
 
 In the hybrid model, the characteristic of the payoff of the IoTSP is similar to the push model, the only difference in the hybrid model is that the payoff of the IoTSP becomes independent of $ba_1$ when $ba_1>2D_{max}/d$ instead of $5D_{max}/d$ (Corollaries~\ref{cor:hybrid_lowad} and \ref{cor:hybrid_highad}). Note that in each model, the IoTSP's payoff is maximized at the maximum value of $ba_1$. Thus, the payoff is maximized at least at $b$ which maximizes\footnote{The above behavior can also be observed in some legacy technologies. For example, Google and Facebook select $b$ in order to maximize the advertisement revenue from each user i.e. $ba_1$.} $ba_1$.

\section{Comparison among Different Models}\label{sec:comparison_models}
We now compare different interaction models to provide insights which model will be preferable to different entities. The significance of the above characterization is immense to the policy maker, and the entities.  For example, if the market is regulated, the policy maker would like to maximize its own objective. If the objective is to increase consumer welfare  it would like to know the model where the demand will be the highest. On the other hand if the object is to incentivize a provider (IoTSP, WSP or CSP) to participate, it needs to know the interaction model which fetches the highest payoff to the provider. If the market is free, the providers and the advertising firms want to adopt the interaction models which give them the highest payoff. Thus, the providers and the advertising firms want to know the interaction models where their payoffs will be the highest. 

We first compare the demand of end-users (Corollary~\ref{cor:demand_compare}) in different interaction models. Subsequently, we compare the payoffs of the IoTSP (Corollary~\ref{cor:iot_compare}), the WSP (Corollary~\ref{cor:wsp_compare}), the CSP (Corollary~\ref{cor:csp_compare}) and the advertisers (Section~\ref{sec:advertiser_compare}) in different interaction models.

We present the result in terms of the chosen $ba_1$. We have discussed in the last section that the IoTSP selects a $b$ which maximizes $ba_1$ in each model. Note that $a_1$ is decreasing function of $b$, thus, $ba_1$ is not maximized at the maximum value of $b$. Table~\ref{table:compare} shows the most preferable model for each entity for high and low values of $ba_1$.
\begin{table}
\caption{\small Most preferable model for various entities in different advertisement revenue regime.}
\begin{center}
\begin{tabular}{| l | c | r |}
\hline
Entities & Sufficiently Low $ba_1$ & Sufficiently High $ba_1$\\ \hline
End-users & Pull & Push\\ \hline
IoTSP & Pull & Pull\\ \hline
WSP & Pull & Push\\ \hline
CSP & Hybrid & Hybrid or Push\\ \hline
Advertisers & Pull & Push\\
\hline
\end{tabular}
\end{center}
\label{table:compare}
\vspace{-0.5cm}
\end{table}
\subsection{The demand of end-users}
\begin{corollary}\label{cor:demand_compare}
\begin{itemize}
\item When $dba_1<D_{max}$, $D_{pull}>D_{push}=D_{hybrid}$.
\item When $dba_1=D_{max}$, $D_{pull}=D_{push}=D_{hybrid}$.
\item When $2D_{max}>= dba_1>D_{max}$, $D_{push}=D_{hybrid}>D_{pull}$
\item When $dba_1>2D_{max}$, $D_{push}>D_{hybrid}>D_{pull}$.
\vspace{-0.1cm}
\end{itemize}
\end{corollary}
At initial stages of deployment of IoT technology, it is expected that $ba_1$ will be small. Thus, if a social planner wants to increase the reach of the IoT technology it may recommend the pull model at initial stages. When $ba_1$ becomes sufficiently high, the social planner then may recommend the push model. 

The demand of end-users decreases with the payment that an end-user incurs. Corollary~\ref{cor:demand_compare} is consistent with the above fact. When $ba_1$ is small (large, resp.) the total payment that an end-user incurs is larger (smaller, resp.) in the push and hybrid models compared to the pull model (Theorems~\ref{thm:push-iot},\ref{thm:pull}, \ref{thm:hybrid}). Thus,  the demand in the push and hybrid model is lower (higher, resp.) compared to the pull model when $ba_1$ is low (high, resp.).  

The total payment that an end-user incurs in the hybrid model is the same as in the push model when $dba_1\leq 2D_{max}$ (Theorems~\ref{thm:push-iot} and \ref{thm:hybrid}), thus, the demand in the push and the hybrid model are the same.  When $dba_1>2D_{max}$, then an end-user has to pay a lower amount in the push model, thus the demand in the push model becomes higher compared to the hybrid model (Theorems~\ref{thm:push-iot} and \ref{thm:hybrid}).  
\begin{figure}
 \includegraphics[width=80mm,height=30mm]{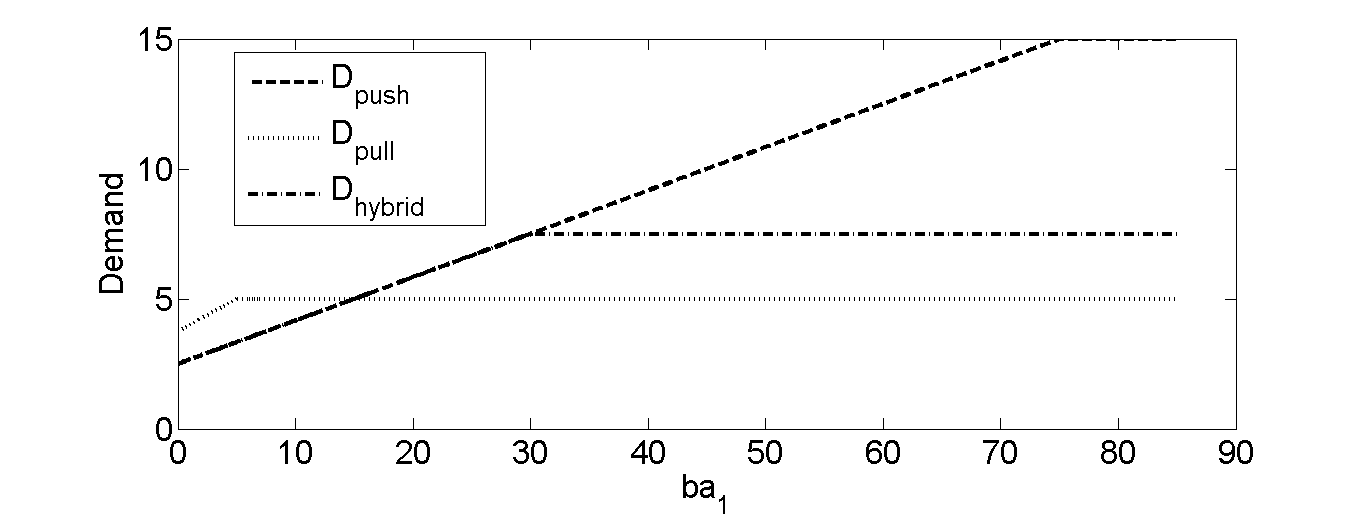}
 \caption{Figure shows the variation of the demand of end-users with $ba_1$ in different interaction models in the same example setting as considered in Fig.~\ref{fig:push_pay}.}
 \label{fig:demand}
 \vspace{-0.4cm}
\end{figure}
 
\begin{figure}
\includegraphics[width=70mm,height=30mm]{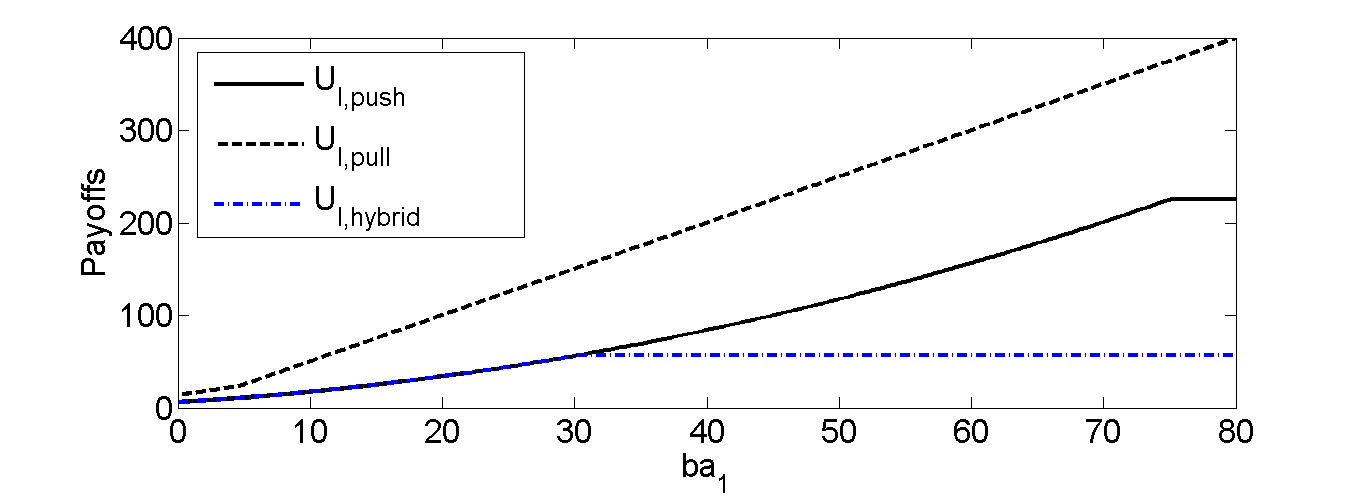}
\caption{Comparison of $U_{I,push}, U_{I,pull}$ and $U_{I,hybrid}$ for the same example setting in fig.~\ref{fig:push_pay}. $U_{I,pull}$ is the highest. $U_{I,push}=U_{I,hybrid}$ when $ba_1\leq 2D_{max}/d=30$, $U_{I,push}>U_{I,hybrid}$ when $ba_1>30$.}
\label{fig:push_pull_hybrid_iotpay}
\vspace{-0.3cm}
\end{figure}
\subsection{The IoTSP\rq{}s payoff}
\begin{corollary}\label{cor:iot_compare}
\item $dba_1\leq 2D_{max}$, $U_{I,pull}>U_{I,push}=U_{I,hybrid}$.
\item $dba_1>2D_{max}$, $U_{I,pull}>U_{I,push}>U_{I,hybrid}$.
\vspace{-0.1cm}
\end{corollary}
Note that the payoff of the IoTSP is the highest in the pull model.  In the pull model, the IoTSP does not have to share its revenue with the CSP or WSP, thus, the payoff is the highest in the pull model in the  low advertisement revenue regime compared to the push and hybrid models (Corollaries~\ref{cor:push_lowad},\ref{cor:pull_lowad}, and \ref{cor:hybrid_lowad}). For high $ba_1$, the payoff of the IoTSP becomes independent of $ba_1$ in the push and hybrid models, but in the pull model, the payoff of the IoTSP always strictly increases with $ba_1$ (Corollaries~\ref{cor:push_highad}, \ref{cor:pull_highad}, and \ref{cor:hybrid_highad}). Thus, the payoff in the pull model is also the highest when $ba_1$ is high. When $dba_1>2D_{max}$, the demand is higher and the IoTSP can gain a higher share of the advertisement revenue in the push model compared to the hybrid model (Corollaries~\ref{cor:push_highad}, and \ref{cor:hybrid_highad}). Hence, the payoff of the IoTSP is higher in the push model compared to the hybrid model when $dba_1>2D_{max}$ (Fig.~\ref{fig:push_pull_hybrid_iotpay}). 
\subsection{Payoff of the WSP}
Recall that $U_{w,worst,push}$ is the worst possible payoff of the WSP in the push model. 
\begin{corollary}\label{cor:wsp_compare}
\begin{itemize}
\item When $0.414D_{max}>dba_1$, $U_{w,pull}>U_{w,push}>U_{w,hybrid}$ . 
\item At $dba_1=0.414D_{max}$, $U_{w,pull}=U_{w,push}>U_{w,hybrid}$.
\item When $D_{max}>dba_1>0.414D_{max}$, $U_{w,push}>U_{w,pull}>U_{w,hybrid}$.
\item When $dba_1=D_{max}$, $U_{w,push}>U_{w,pull}=U_{w,hybrid}$.
\item When $5D_{max}>dba_1>D_{max}$, $U_{w,push}>U_{w,hybrid}>U_{w,pull}$.
\item When $dba_1\geq 5D_{max}$, $U_{w,worst,push}>U_{w,hybrid}>U_{w,pull}$.
\end{itemize}
\vspace{-0.1cm}
\end{corollary}
Though the WSP is a leader in the push model, the payoff of the WSP is lower in the push model compared to the pull model when $ba_1$ is small. This is because, when $ba_1$ is small the demand is low which fetches a lower payoff to the WSP in the push model compared to the pull model. With increase in $ba_1$, the payoff of the WSP increases at a higher rate in the push model compared to the pull model (Corollaries~\ref{cor:push_lowad}, and \ref{cor:pull_lowad}).  Thus, the WSP\rq{}s payoff becomes higher in the push model compared to the pull model in the high advertisement revenue regime.
The price set by the WSP is always higher in the push model compared to the hybrid model (Theorems~\ref{thm:push}, and \ref{thm:hybrid}), thus, the payoff of the WSP is always higher in the push model compared to the hybrid model.   

Note that the WSP quotes its price to the end users both in the pull model and hybrid model. However, when $ba_1$ is small (high, resp.) the demand in the hybrid model is small (high, resp.) compared to the pull model, thus, the payoff of the WSP is  higher (lower, resp.) in the pull model.    Note that when $ba_1\geq 5D_{max}/d$, then the payoff of the WSP is not unique in the push model as the equilibrium is not unique (Theorem~\ref{thm:push}) and depends on the equilibrium selected. However, from Corollaries~\ref{cor:push_highad}, \ref{cor:pull_highad}, and \ref{cor:hybrid_highad}, the worst case payoff of the WSP in the push model is the highest. Hence, {\em irrespective of the equilibrium selected, the payoff of the WSP is the highest in the push model when $dba_1\geq 5D_{max}$.}
\subsection{The CSP\rq{}s payoff}
 In the push model, since the SPNE may not be unique when $dba_1>5D_{max}$ (Theorem~\ref{thm:push}), the payoff of the CSP is not unique and will depend on the choice of an equilibrium. Thus, we compare with the worst possible payoff ($U_{c,worst,push}$) and the best possible payoff ($U_{c,best,push}$) of the CSP in the push model.
\begin{corollary}\label{cor:csp_compare}
\begin{itemize}
\item When $dba_1<0.414D_{max}$, $U_{c,hybrid}>U_{c,pull}>U_{c,push}$. 
\item At $dba_1=0.414D_{max}$, $U_{c,hybrid}>U_{c,pull}=U_{c,push}$.
\item When $5.5D_{max}> dba_1>0.414D_{max}$, $U_{c,hybrid}>U_{c,push}>U_{c,pull}$.
\item When $dba_1\geq 5.5 D_{max}$, $U_{c,hybrid}>U_{c,worst,push}>U_{c,pull}$.
\item When $dba_1=5.5D_{max}$, $U_{c,hybrid}=U_{c,best,push}>U_{c,pull}$.
\item When $dba_1>5.5D_{max}$, $U_{c,best,push}>U_{c,hybrid}>U_{c,pull}$.
\end{itemize}
\end{corollary}
 Since the CSP along with the WSP are leaders of the game in the push model, the comparison between the payoffs of the CSP in the push and pull model is similar to that of the WSP.
 However, since the CSP is the only leader in the hybrid model,  it can select higher prices and still it can attain a higher payoff in the hybrid model compared to the push and pull model for small values of $ba_1$(Corollaries~\ref{cor:push_lowad}, \ref{cor:pull_lowad}, and \ref{cor:hybrid_lowad}).  For high $ba_1$, the  CSP\rq{}s payoff increases linearly with $ba_1$ in the hybrid model, whereas $U_{c,worst,push}$ (Corollary~\ref{cor:push_highad}) and $U_{c,pull}$ (Corollary~\ref{cor:pull_highad}) become independent of $ba_1$. Thus, $U_{c,hybrid}$ is higher compared to $U_{c,pull}$ and $U_{c,worst,push}$ for high $ba_1$. However, the best possible payoff of the CSP in the push model is higher  compared to the payoff in the hybrid model when $dba_1>5.5D_{max}$. This is because the best possible payoff of the CSP in the push model also increases linearly with the advertisement revenue (Corollary~\ref{cor:push_highad}) and  the demand in the push model is higher compared to the hybrid model when $dba_1>5.5D_{max}$ (Corollary~\ref{cor:demand_compare}) . 
 
\subsection{Expected payoff of the advertisers}\label{sec:advertiser_compare}
Recall from Section~\ref{game:IoTSP and advertisers} that each advertising firm decides whether to participate in advertising or not. Advertising firm $i$ gets an expected revenue $r_i$ from each end-user if it decides to advertise. If the advertising firm does not participate, then it will get a revenue of $0$ from each end-user.  The advertising firm also has to pay $b$ if it decides to participate. Thus, the total expected payoff of  advertising firm $i$ is--
 \begin{align}\label{eq:utility_advertiser}
 &(r_i-b)*D\quad \mbox{if firm i participates }\nonumber\\
 & 0 \quad \mbox{otherwise}
  \end{align}
  where $D$ is the total demand of the end-users.  Hence, the expected payoff of  all participating advertising firms  increases with the increase in the demand of end-users.

 Note from Corollary~\ref{cor:demand_compare} that the demand of end-users depend on $ba_1$, the total advertisement revenue for each user. The demand will impact the revenue of participating advertising firms and thus, it should impact the number of participating firms $a_1$. However, we consider that $a_1$ is not a strategic decision and is a function of only $b$ (recall (\ref{eq:advertising})). We now justify it.  In practice, each advertising firm is an independent entity and there is no coordination among different advertising firms.  Each advertising  firm only maximizes its own payoff and thus, it decides to participate based on its own expected revenue from each end-user and $b$. Thus, the number of participating advertising firms $a_1$ is a function of  only $b$.There is no central controller which coordinates among the advertising firms to select $a_1$ to maximize the sum of the payoffs of the advertising firms.  Hence, $a_1$ is not a strategic decision.  
 
Recall that the IoTSP selects $b$ which maximizes $ba_1$.  For a given $b$, from (\ref{eq:utility_advertiser}) the total expected payoff of each participating firm\footnote{A firm will participate only if $r_i\geq b$.} is a non-decreasing function of the demand $D$. Now from  Corollary~\ref{cor:demand_compare}, when the chosen $ba_1$ is low (high, resp.)  then the total demand is the highest (lowest, resp.) in the pull model. Thus, when chosen $ba_1$ is low (high,resp.) , the payoff of each participating advertising firm is the highest (lowest, resp.) in the pull model.  When $ba_1>2D_{max}/d$, the demand and thus,  total expected payoff of a participating advertiser is higher in the push model compared to the hybrid otherwise, the demand and thus, the  total expected payoffs of a participating advertiser are equal in the push and hybrid model (by Corollary~\ref{cor:demand_compare}). 
 \begin{figure}
 \includegraphics[width=80mm,height=30mm]{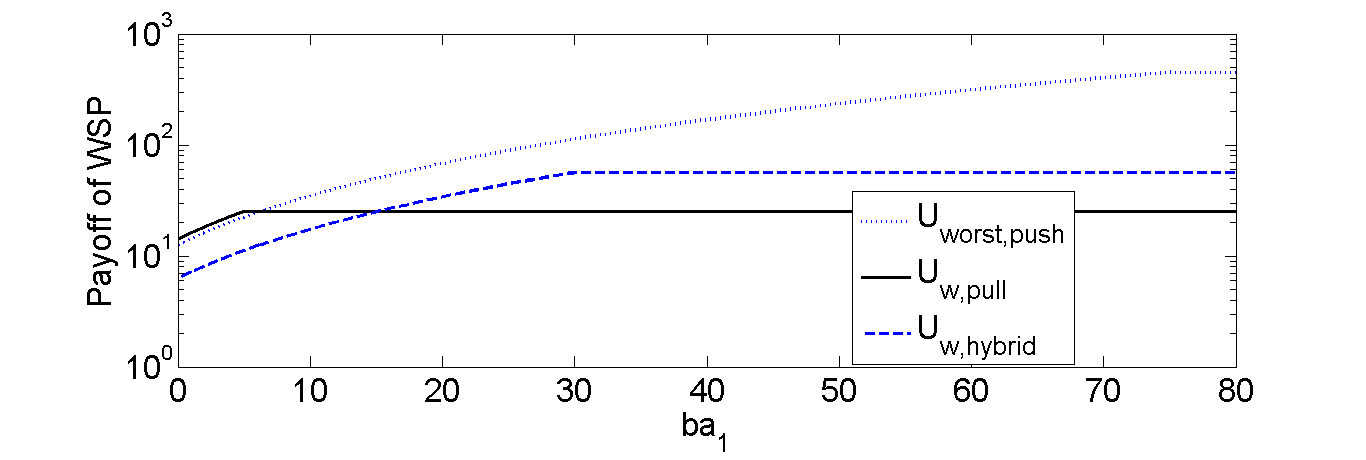}
 \caption{\small Comparison of $U_{w,worst,push}, U_{w,pull}$, and $U_{w,hybrid}$ for the same example setting in Fig.~\ref{fig:push_pay}. For $dba_1\leq 5D_{max}=75$, the WSP\rq{}s payoff is unique in the push model, thus, $U_{w,worst,push}=U_{w,push}$.}
 \label{fig:wsp_payoff}
 \vspace{-0.4cm}
\end{figure}
 \begin{figure}
\includegraphics[width=90mm,height=40mm]{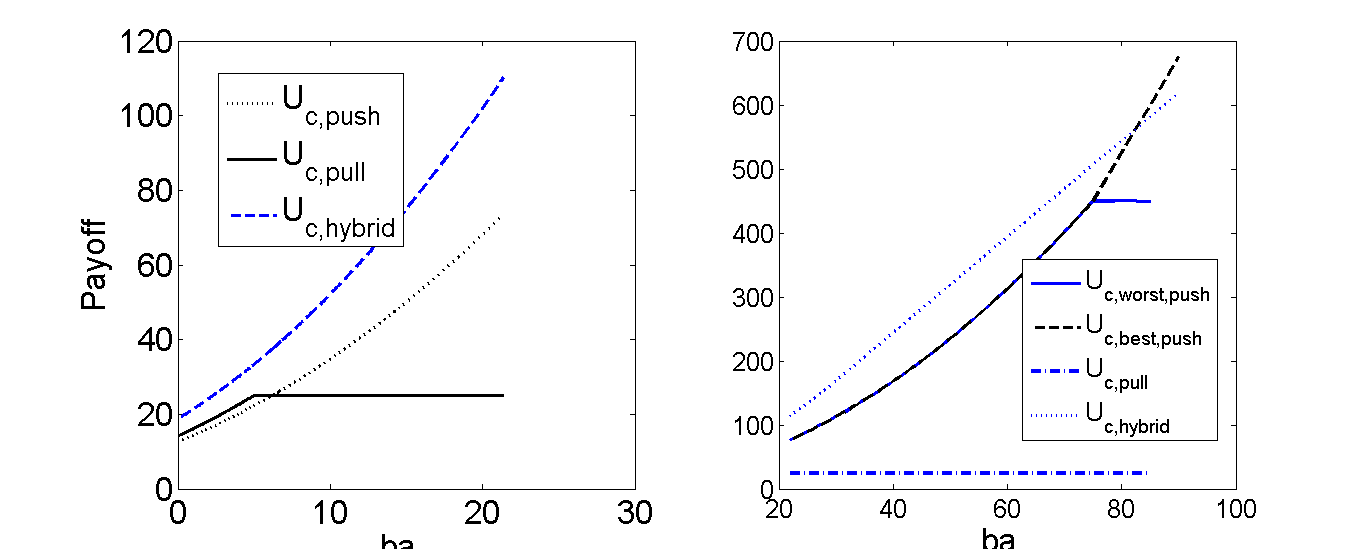}
\caption{\small Figure in the left side shows the CSP\rq{}s payoff in different interaction models for small values of $ba_1$ ($ba_1\leq 22$) for the same example setting in Fig.~\ref{fig:push_pay}. The figure in the right hand side shows the CSP\rq{}s payoff in different interaction models for higher values of $ba_1$ ($ba_1>22$) for the same example setting in Fig.~\ref{fig:push_pay}. Since when $dba_1>5D_{max}=75$, the payoff of the IoTSP is not unique in the push model, we plot the best and worst possible payoffs.}
 \label{fig:csp_payoff}
 \vspace{-0.5cm}
\end{figure}
 
\section{Numerical Results}
We now numerically evaluate the demand of end-users, and the payoff of the providers as a function of $ba_1$ in an example scenario.

 Fig.~\ref{fig:demand} reveals that the demand of end-users is the highest (lowest, resp.) in the pull model for low (high, resp.) advertisement revenue regime which we also found in Corollary~\ref{cor:demand_compare}. 
 As Corollary~\ref{cor:demand_compare} suggests the demand in the push model is higher (the same, resp.) when $ba_1>2D_{max}/d$ ($ba_1\leq 2D_{max}/d$, resp.) compared to the hybrid model. Thus, the demand is the highest in the push model  when $dba_1>2D_{max}$. 

Fig.~\ref{fig:push_pull_hybrid_iotpay} shows that the payoff of the IoTSP is the highest in the pull model (Corollary~\ref{cor:iot_compare}). The payoff of the IoTSP is higher in the push model compared to the hybrid model when $ba_1> 2D_{max}/d$ otherwise it is equal as we found in Corollary~\ref{cor:iot_compare}. Fig.~\ref{fig:push_pull_hybrid_iotpay}  reveals that the difference between the payoffs in the push (or, hybrid) and pull model strictly increases with $ba_1$.

Fig.~\ref{fig:wsp_payoff} shows the variation of the WSP\rq{}s payoff with $ba_1$ in different interaction models. 
Since $U_{w,push}$ is not unique when $dba_1>5D_{max}$, we plot the worst case payoff of the WSP. $U_{w,push}$ is unique when $dba_1\leq 5D_{max}$, hence, the worst case payoff in this region is equal to $U_{w,push}$. Note that even the worst case payoff of the WSP in the push model is the highest  when $dba_1>5D_{max}$ as suggested in Corollary~\ref{cor:wsp_compare}. Thus, irrespective of the equilibrium selected in the push model for $ba_1>5D_{max}/d$, the payoff of the WSP is the highest in the push model. However, the payoff of the WSP is the highest in the pull model for small values of $ba_1$. The payoff of the WSP is higher (lower, resp.) in the hybrid model compared to the pull model for higher (lower, resp.) values of $ba_1$ as suggested in Corollary~\ref{cor:wsp_compare}.

Fig.~\ref{fig:csp_payoff} shows the variation of the CSP\rq{}s payoff with $ba_1$ in different interaction  models. It shows that the payoff of the CSP is the highest in the hybrid model for low values of $ba_1$ (Corollary~\ref{cor:csp_compare}). The payoff of the CSP is higher (lower, resp.) in the pull model compared to the push model for small (large, resp.) values of $ba_1$. When $dba_1\leq 5D_{max}$, the payoff the CSP is unique in the push model, thus, the worst and best possible payoffs of the CSP  are identical in the push model (Corollary~\ref{cor:push_lowad}). However, the worst and best possible payoffs of the CSP are not identical when $dba_1>5D_{max}$ in the push model (Corollary~\ref{cor:push_highad}).  The best possible payoff of the CSP in the push model exceeds the payoff in the hybrid model when $ba_1>5.5D_{max}/d$. However, the worst possible payoff of the CSP in the push model is lower than the payoff in the hybrid model as we found in Corollary~\ref{cor:csp_compare}. 

\section{Future Work}

  Generalization of our framework to account for an oligopolistic setting is a work for the future. In future we also consider the non-linear demand response models. The bandwidth of the WSP is a scarce resource. Due to the proliferation of the IoT  a large amount of bandwidth is required. In the future, the WSP may need to procure additional bandwidth from the secondary market in exchange of some fee. Thus, the WSP also has to decide over additional bandwidth to be procured apart from the price it will charge the IoTSP or the end-users. The characterization of the equilibrium pricing strategies in the above setting  is also a work for the future. The framework that we have developed will provide a basis for  solving the above problems.

\bibliographystyle{ieeetran}
\appendix
We show the results stated in Section~\ref{sec:push} in Appendix~\ref{proof:push}, results stated in Section~\ref{sec:pull} in Appendix~\ref{proof:pull}, results stated in Section~\ref{sec:hybrid} in Appendix~\ref{proof:hybrid}, and results stated in Section~\ref{sec:comparison_models} in Appendix~\ref{proof:compare}.
\subsection{Proofs of Results in Section~\ref{sec:push}}\label{proof:push}
We first prove Lemma~\ref{lm:push}; subsequently we prove Theorem~\ref{thm:push}. Finally, we  show Corollaries~\ref{cor:push_lowad} and \ref{cor:push_highad}.

We use the following result throughout this section. Note from  (\ref{demand}) and (\ref{utilitypush}) that
\begin{align}\label{utilityiot}
U_{I,push}= p_I(D_{max}-dp_I)+ba_1(D_{max}-dp_I)-\nonumber\\
p_w(\alpha+a_1)(D_{max}-dp_I)-p_c(\beta+a_2)(D_{max}-dp_I)
\end{align}
Now we are ready to show Lemma~\ref{lm:push}.

\textit{Proof of Lemma~\ref{lm:push}:}
From the first order condition and (\ref{utilityiot}), we must have at NE
\begin{align}\label{iotpricepush1}
p_I^{*}=\dfrac{D_{max}}{2d}-\dfrac{ba_1}{2}+\dfrac{p_w(\alpha+a_1)}{2}+\dfrac{p_c(\beta+a_2)}{2}
\end{align}
Note that when IoTSP selects price $p_I$ it knows $p_w$, $p_c$.  Since the $U_{I,push}$ is strictly concave in $p_I$ by (\ref{utilityiot}) thus the first order condition is both necessary and sufficient for optimality. Thus, (\ref{iotpricepush1}) is also the optimal price for IoTSP if it is non-negative.  If the right hand side of (\ref{iotpricepush1}) is negative then $p_I^{*}=0$. Hence, the result follows.\qed

\textit{Proof of Theorem~\ref{thm:push}:}
 We prove the theorem by considering three cases:\\
 (i) We first show that  an NE can not arise in the case where $\dfrac{p_w(\alpha+a_1)}{2}+\dfrac{p_c(\beta+a_2)}{2}<\dfrac{ba_1}{2}-\dfrac{D_{max}}{2d}$  since in this case either the WSP or the CSP can increases its price without decreasing the demand $D_{push}$ and thus, can gain higher payoff by unilaterally deviating from the strategy profile (Step 1).

(ii) We next consider the case when $\dfrac{p_w(\alpha+a_1)}{2}+\dfrac{p_c(\beta+a_2)}{2}>\dfrac{ba_1}{2}-\dfrac{D_{max}}{2d}$ (Step 2). In this case we first obtain the expression of $D_{push}$, $U_{w,push}$, and $U_{c,push}$ from (\ref{iotpricepush1}) (Step 2a). Subsequently, we characterize the equilibrium strategy profile from the first order condition since the payoffs are strictly concave in this case (Step 2b). 
 
 (iii) Finally we consider the case when $\dfrac{p_w(\alpha+a_1)}{2}+\dfrac{p_c(\beta+a_2)}{2}=\dfrac{ba_1}{2}-\dfrac{D_{max}}{2d}$ (Step 3). We obtain the expression of $D_{push}$ and $U_{w,push}, U_{c,push}$ using (\ref{iotpricepush1}) (Step 3a). The first order condition is no longer sufficient in this case since the payoffs are not strictly concave, nevertheless we characterize the equilibrium prices and the necessary and sufficient conditions required for the equilibrium (Step 3b).

The detailed proof is given below:

{\em Step 1}: If $\dfrac{p_w(\alpha+a_1)}{2}+\dfrac{p_c(\beta+a_2)}{2}<\dfrac{ba_1}{2}-\dfrac{D_{max}}{2d}$
then $p_I^{*}=0$ by (\ref{iotpricepush}) and thus from (\ref{demand}) the demand becomes
\begin{align}
D_{push}=D_{max}-dp_I=D_{max}
\end{align}
Thus, CSP\rq{}s payoff is 
\begin{align}
p_c(\beta+a_2)D_{max}\nonumber
\end{align}
which is a strictly increasing function in $p_c$. Since $p_c(\beta+a_2)<ba_1-D_{max}/d-p_w(\alpha+a_1)$,thus, we can always find a small enough $\epsilon>0$ such that  $(p_c+\epsilon)(\beta+a_2)<ba_1-D_{max}/d-p_w(\alpha+a_1)$, but the payoff at $p_c+\epsilon$ is  $(p_c+\epsilon)(\beta+a_2)D_{max}$ which is strictly higher than the payoff at price $p_c$ which contradicts the fact that $p_c$ is an NE. Thus, NE can not arise in this case.

{\em Step 2}: We now consider the case 
\begin{align}\label{pushcondition}
\dfrac{p_w(\alpha+a_1)}{2}+\dfrac{p_c(\beta+a_2)}{2}> \dfrac{ba_1}{2}-\dfrac{D_{max}}{2d}
\end{align}
{\em Step 2a}: Replacing the value of $p_I^{*}$ from (\ref{iotpricepush}) in (\ref{demand}) we get
\begin{align}\label{demandi}
D_{push}=\dfrac{D_{max}}{2}-d\left(-\dfrac{ba_1}{2}+\dfrac{p_w(\alpha+a_1)}{2}+\dfrac{p_c(\beta+a_2)}{2}\right)
\end{align}
Using (\ref{demandi}), (\ref{utilitywsppush}), and (\ref{utilitycsppush}) we can easily show that 
\begin{align}\label{equation_w}
U_{w,push}& =p_w(\alpha+a_1)(\dfrac{D_{max}}{2}-\nonumber\\
& d(-\dfrac{ba_1}{2}+\dfrac{p_w(\alpha+a_1)}{2}+\dfrac{p_c(\beta+a_2)}{2}))\\
U_{c,push}& =p_c(\beta+a_2)(\dfrac{D_{max}}{2}\nonumber\\& -d(-\dfrac{ba_1}{2}+\dfrac{p_w(\alpha+a_1)}{2}+\dfrac{p_c(\beta+a_2)}{2}))
\end{align}
{\em Step 2b}:
The payoff functions of the WSP and CSP are strictly concave in $p_w$ and $p_c$ respectively in this case. Thus, the first order condition is also the sufficient one if they satisfy the condition in (\ref{pushcondition}).
Thus, at NE pricing strategy $(p_w^{*},p_c^{*})$, we must have from the first order condition
\begin{align}
p_w^{*}=\dfrac{D_{max}}{2d(\alpha+a_1)}-\dfrac{p_c^{*}(\beta+a_2)}{2(\alpha+a_1)}+\dfrac{ba_1}{2(\alpha+a_1)}\nonumber\\
p_c^{*}=\dfrac{D_{max}}{2d(\beta+a_2)}-\dfrac{p_w^{*}(\alpha+a_1)}{2(\beta+a_2)}+\dfrac{ba_1}{2(\beta+a_2)}\nonumber
\end{align}
given that $(p_w^{*},p_c^{*})$ satisfy condition in (\ref{pushcondition}). Further simplification leads to
\begin{eqnarray}
p_w^{*}=\dfrac{D_{max}}{3d(\alpha+a_1)}+\dfrac{ba_1}{3(\alpha+a_1)}\label{n1}\\
p_c^{*}=\dfrac{D_{max}}{3d(\beta+a_2)}+\dfrac{ba_1}{3(\beta+a_2)}\label{n2}
\end{eqnarray}
Note that condition (\ref{pushcondition}) is satisfied when
\begin{align}
p_c^{*}(\beta+a_2)+p_w^{*}(\alpha+a_1)> ba_1-D_{max}/d\nonumber\\
ba_1< 5D_{max}/d
\end{align}
Thus, the NE strategy profile exists in this case only when $dba_1<5D_{max}$.

{\em Step 3}: We now consider the case
\begin{align}\label{pushcondition1}
\dfrac{p_w(\alpha+a_1)}{2}+\dfrac{p_c(\beta+a_2)}{2}= \dfrac{ba_1}{2}-\dfrac{D_{max}}{2d}
\end{align}
{\em Step 3a}:Here, $p_I^{*}=0$ by (\ref{iotpricepush}). Thus, the demand becomes
\begin{align}
D=D_{max}-dp^{*}_I=D_{max}
\end{align}
So, WSP\rq{}s payoff and CSP\rq{}s payoff is
\begin{align}\label{eq:monopoly_w}
U_{w,push}=p_w(\alpha+a_1)D_{max}\nonumber\\
U_{c,push}=p_c(\beta+a_2)D_{max}
\end{align}
{\em Step 3b}:
Now, we will find out the conditions for which $(p_w,p_c)$ is an NE in this case. To this end, we have to rule out any profitable unilateral deviation by either WSP or CSP.

We first show that when the WSP selects a price less than $p_w$, then the WSP\rq{}s payoff will be strictly less than the payoff at $p_w$ (Case iii.a). Subsequently, we find the conditions which ensure that the WSP will not have any profitable deviation when it selects a price higher than $p_w$ (Case iii.b).  Due to the symmetry, the analysis for the CSP will be similar.

\textit{Case iii.a}: If the WSP selects a lower price $x$, then by condition in (\ref{pushcondition1}), $x(\alpha+a_1)/2+p_c(\beta+a_2)/2< ba_1/2-D_{max}/(2d)$ i.e. it satisfies the condition in case i. Thus, $p_I^{*}=0$. Thus, the WSP\rq{}s payoff becomes $x(\alpha+a_1)D_{max}$ which is strictly less than $p_w(\alpha+a_1)$ since $x<p_w$. Thus, the WSP has no incentive to lower its price.

\textit{Case iii b}: Now, we obtain the condition under which the WSP will not have any incentive to select a price larger than $p_w$.  Suppose WSP selects price $x>p_w$, thus,  $x(\alpha+a_1)+p_c(\beta+a_2)>ba_1-D_{max}$; i.e it satisfies the condition in case ii. Thus, from (\ref{equation_w}) the WSP\rq{}s payoff at the price $x$ is
\begin{align}
x(\alpha+a_1)\left(\dfrac{D_{max}}{2}-d\left(-\dfrac{ba_1}{2}+\dfrac{x(\alpha+a_1)}{2}
+\dfrac{p_c(\beta+a_2)}{2}\right)\right)\nonumber
\end{align}
Now, using (\ref{pushcondition1}) in place of $p_c(\beta+a_2)$, the above expression can be written as
\begin{align}\label{payoffatx}
x(\alpha+a_1)(D_{max}/2\nonumber\\
-d(-\dfrac{D_{max}}{2d}-p_w(\alpha+a_1)/2+x(\alpha+a_1)/2))
\end{align}
Hence, there will be no profitable unilateral deviation if payoff at $p_w$ is greater than or equal to the payoff at $x$, thus from (\ref{eq:monopoly_w}) and (\ref{payoffatx}), we must have
\begin{align}
& (\alpha+a_1)(p_w-x)(D_{max}-dx(\alpha+a_1)/2)\geq 0
\end{align}
Since $x>p_w$, thus the above condition will be satisfied only when $2D_{max}/d\leq x(\alpha+a_1)$. If $2D_{max}/d>p_w(\alpha+a_1)$, then, we can find $x>p_w$ such that $x(\alpha+a_1)<2D_{max}/d$. Hence, for no profitable unilateral deviation for the WSP, we must have $2D_{max}/d\leq p_w(\alpha+a_1)$. Similarly, we can also show that for no profitable unilateral deviation for the CSP, we must have $2D_{max}/d\leq p_c(\beta+a_2)$. Now if $p_w(\alpha+a_1)>ba_1-3D_{max}/d$, then by (\ref{pushcondition1})
\begin{align}
p_c(\beta+a_2)<2D_{max}/d\nonumber
\end{align}
Thus, we must have  $p_w(\alpha+a_1)\leq ba_1-3D_{max}/d$ .

Again, since $p_w(\alpha+a_1)\geq 2D_{max}/d$, thus, for feasible $p_w$ we must have $2D_{max}/d\geq ba_1-3D_{max}/d$ i.e. $dba_1\geq 5D_{max}$ i.e. this case can arise only when $dba_1\geq 5D_{max}$. Similarly, in order to satisfy the condition that $p_w(\alpha+a_1)\geq 2D_{max}/d$ and (\ref{pushcondition1}), we must have $p_c(\alpha+a_1)\leq ba_1-3D_{max}/d$. Since $p_c(\beta+a_2)\geq 2D_{max}/d$, thus, similarly $p_c$ is feasible only when $dba_1\geq 5D_{max}$.

Thus, from case iii.b an NE can exist in this case only when $dba_1\geq 5D_{max}$ and the NE strategy  $(p_w^{*},p_c^{*})$ must be of the following from
\begin{align}
p_w^{*}(\alpha+a_1)\in [2D_{max}/d,ba_1-3D_{max}/d]\nonumber\\
p_c^{*}(\beta+a_2)\in [2D_{max}/d,ba_1-3D_{max}/d]\nonumber\\
\text{such that } p_c^{*}(\beta+a_2)+p_w^{*}(\alpha+a_1)=ba_1-D_{max}/d\nonumber
\end{align}
The last equality comes from (\ref{pushcondition1}). Note that when $dba_1\geq 5D_{max}$, there may exist multiple  NE pricing strategy in this sub-game. However, when $dba_1=5D_{max}$ then, the NE strategy becomes unique, $p_w^{*}(\alpha+a_1)=p_c^{*}(\beta+a_1)=2D_{max}/d$.

Hence, from case ii,  when $dba_1<5D_{max}$, the NE strategy must be of the form given in (\ref{eq:wsppricelowad})-(\ref{eq:csppricelowad}) and from case iii, when $dba_1\geq 5D_{max}$, then NE strategy profile must be of the form given in (\ref{eq:wsppricehighad})-(\ref{eq:csppricehighad}). \qed

\textit{Proof of Corollary~\ref{cor:push_lowad}}
We first obtain the expression of $D_{push}$ from (\ref{iotpricepush}) and (\ref{demand}) when $dba_1<5D_{max}$.  Using the value of $D_{push}$ and Theorems~\ref{thm:push} and \ref{thm:push-iot} we subsequently obtain the values of $U_{I,push}, U_{w,push}$ and $U_{c,push}$. 

From (\ref{iotpricepush}) and (\ref{demand}) we obtain
\begin{align}\label{demand_lowad}
D_{push}=D_{max}-dp_I=\dfrac{D_{max}}{6}+d\dfrac{ba_1}{6}
\end{align}
We now obtain the expression of $U_{I,push}$ when $dba_1<5D_{max}$. From (\ref{utilitypush}), (\ref{iotpricepush}), (\ref{eq:wsppricelowad}),(\ref{eq:csppricelowad}) and (\ref{demand_lowad}) we obtain 
\begin{align}\label{utilitypushiotspfinal}
& U_{I,push}=(\dfrac{5D_{max}}{6d}-\dfrac{ba_1}{6})(\dfrac{D_{max}}{6}+d\dfrac{ba_1}{6})\nonumber\\& +ba_1(\dfrac{D_{max}}{6}+d\dfrac{ba_1}{6})\nonumber\\& -(\dfrac{D_{max}}{3d(\alpha+a_1)}+\dfrac{ba_1}{3(\alpha+a_1)})(\alpha+a_1)(\dfrac{D_{max}}{6}+d\dfrac{ba_1}{6})\nonumber\\& -(\dfrac{D_{max}}{3d(\beta+a_2)}+\dfrac{ba_1}{3(\beta+a_2)})(\beta+a_2)(\dfrac{D_{max}}{6}+d\dfrac{ba_1}{6})\nonumber\\
& U_{I,push}=(\dfrac{D_{max}}{6d}+\dfrac{ba_1}{6})(\dfrac{D_{max}}{6}+d\dfrac{ba_1}{6})\nonumber\\
& U_{I,push}=d(\dfrac{D_{max}}{6d}+\dfrac{ba_1}{6})^2
\end{align}
Now, we obtain the expression of $U_{w,push}$ when $dba_1<5D_{max}$. We obtain from (\ref{utilitywsppush}),(\ref{eq:wsppricelowad}) and (\ref{demand_lowad})
\begin{align}\label{utilitypushwspfinal}
U_{w,push}& =p_w^{*}(\alpha+a_1)(\dfrac{D_{max}}{6}+d\dfrac{ba_1}{6})\nonumber\\
& =(\dfrac{D_{max}}{3d}+\dfrac{ba_1}{3})(\dfrac{D_{max}}{6}+d\dfrac{ba_1}{6})\nonumber\\
& =2d\left(\dfrac{D_{max}}{6d}+\dfrac{ba_1}{6}\right)^2
\end{align}
Similarly from (\ref{utilitycsppush}), (\ref{eq:csppricelowad}) and (\ref{demand_lowad}) we obtain
\begin{align}\label{utilitypushcspfinal}
U_{c,push}& =2d(\dfrac{D_{max}}{6d}+\dfrac{ba_1}{6})^2
\end{align}
Corollary~\ref{cor:push_lowad} readily follows from (\ref{demand_lowad})-(\ref{utilitypushcspfinal}).\qed

\textit{Proof of Corollary~\ref{cor:push_highad}}
We first obtain the expression of the equilibrium value of $D_{push}$ by plugging in the value of equilibrium $p_I^{*}$ from (\ref{iotpricepush}) when $dba_1\geq 5D_{max}$. We obtain the expression of $U_{I,push}$ using the equilibrium value of demand and (\ref{eq:highadcondition}). The payoffs of the WSP and CSP are not unique in the push model when $dba_1\geq 5D_{max}$ since the equilibrium is not unique (Theorem~\ref{thm:push-iot}). We obtain the worst case and best possible payoffs of the WSP and CSP from Theorem~\ref{thm:push} and the equilibrium demand when $dba_1\geq 5D_{max}$. Finally, we show that the sum of the payoffs of the WSP and CSP are unique and obtain the expression.

When $dba_1\geq 5D_{max}$, then
$p_I^{*}=0$ by (\ref{iotpricepush}), the equilibrium demand is thus-
\begin{align}\label{eq:demandhighad}
D_{push}=& D_{max}-dp_I=D_{max}
\end{align}
Now, we obtain the expression of the $U_{I,push}$. Note that $p_w^{*}(\alpha+a_1)+p_c^{*}(\beta+a_2)$ is equal to $ba_1-D_{max}/d$ by (\ref{eq:highadcondition}) when $dba_1\geq 5D_{max}$, thus from (\ref{utilitypush}) and (\ref{iotpricepush}) we obtain
\begin{align}
U_{I,push}
=& 0*D_{push}+ba_1D_{push}-\nonumber\\
& (p_w^{*}(\alpha+a_1)+p_c^{*}(\beta+a_2))D_{push}\nonumber\\
=& \dfrac{D_{max}^2}{d}\quad (\text{from } (\ref{eq:demandhighad}))\label{utilityiotsphighad1}
\end{align}
We now obtain the worst and the best possible payoffs of the WSP and CSP. Note from (\ref{eq:wsppricehighad}) and (\ref{eq:csppricehighad}) that the lowest possible values for $p_w(\alpha+a_1)$ and $p_c(\beta+a_2)$ are $2D_{max}/d$. Thus, from (\ref{eq:demandhighad}) the worst possible payoffs for WSP and CSP are respectively-
\begin{eqnarray}
\tilde{U}_{w,push}=& \dfrac{2D_{max}^2}{d} \quad \label{eq:utilitywsppushhighad}\\
\tilde{U}_{c,push}=& \dfrac{2D_{max}^2}{d}\label{eq:utilitycsppushhighad}
\end{eqnarray}
Similarly note from (\ref{eq:wsppricehighad}) and (\ref{eq:csppricehighad}) the the highest possible values of $p_w(\alpha+a_1)$ and $p_c(\beta+a_2)$ are $ba_1-3D_{max}/d$.  Thus, from (\ref{eq:demandhighad}) the best possible payoffs of the WSP and CSP are respectively
\begin{eqnarray}
U_{w,best,push}=& (ba_1-3D_{max}/d)D_{max}\quad \label{eq:utilitywsppushhighadbest}\\
U_{c,best,push}=& (ba_1-3D_{max}/d)D_{max}\quad \label{eq:utilitycsppushhighadbest}
\end{eqnarray}
From (\ref{eq:highadcondition}) and (\ref{eq:demandhighad}) we obtain
\begin{align}\label{eq:paywspcsp}
(p_w^{*}(\alpha+a_1)+p_c(\beta+a_2))D_{push}=ba_1D_{max}-\dfrac{D_{max}^2}{d}
\end{align}
Corollary~\ref{cor:push_highad} readily follows from (\ref{eq:demandhighad})-(\ref{eq:paywspcsp}).\qed

\subsection{Proofs of Results in Section~\ref{sec:pull}}\label{proof:pull}
We first prove Theorem~\ref{thm:pull}. Subsequently we prove Corollaries~\ref{cor:pull_lowad} and \ref{cor:pull_highad}.. 

\textit{Proof of Theorem~\ref{thm:pull}}
We first obtain the expressions of $U_{I,pull}, U_{w,pull}$ and $U_{c,pull}$ using  (\ref{demandpull}), (\ref{utilitypull}), (\ref{utilitypullwsp}) and (\ref{utilitypullcsp})   (Step 1). The payoffs are strictly concave in their respective decision variables, hence, the first order conditions are necessary and sufficient. From the first order condition, we show that only $p_I^{*}$ can be $0$ (Step 2). Subsequently,  we characterize the equilibrium prices in two different cases when $p_I^{*}>0$ and when $p_{I}^{*}=0$ respectively (Steps 3 and 4). The detailed proof is given below:

{\em Step 1}
First, we obtain the expression of $U_{I,pull}$. Replacing (\ref{demandpull}) in (\ref{utilitypull}) we obtain
\begin{align}
U_{I,pull}=p_I(D_{max}-d(p_I+p_w(\alpha+a_1)+p_c(\beta+a_2)))\nonumber\\
+ba_1(D_{max}-d(p_I+p_w(\alpha+a_1)+p_c(\beta+a_2)))\nonumber
\end{align}
Now, we obtain the expressions of $U_{w,pull}$ and $U_{c,pull}$. Replacing (\ref{demandpull}) in (\ref{utilitypullwsp}) and (\ref{utilitypullcsp}) we also obtain
\begin{align}
U_{w,pull}=p_w(\alpha+a_1)(D_{max}\nonumber\\
-d(p_I+p_w(\alpha+a_1)+p_c(\beta+a_2)))\nonumber\\
U_{c,pull}=p_c(\beta+a_2)(D_{max}\nonumber\\-d(p_I+p_w(\alpha+a_1)+p_c(\beta+a_2)))\nonumber
\end{align}
{\em Step 2}:  From the first order condition
\begin{eqnarray}
p^{*}_I=\max\{\dfrac{D_{max}}{2d}-\dfrac{p_w^{*}(\alpha+a_1)}{2}-\dfrac{p^{*}_c(\beta+a_2)}{2}-\dfrac{ba_1}{2},0\}\label{iotspprice}\\
p^{*}_w=\max\{\dfrac{D_{max}}{2d(\alpha+a_1)}-\dfrac{p^{*}_I}{2(\alpha+a_1)}-\dfrac{p^{*}_c(\beta+a_2)}{2(\alpha+a_1)},0\}\nonumber\\
p^{*}_c=\max\{\dfrac{D_{max}}{2d(\beta+a_2)}-\dfrac{p^{*}_I}{2(\beta+a_2)}-\dfrac{p^{*}_w(\alpha+a_1)}{2(\beta+a_2)},0\}\nonumber
\end{eqnarray}
Note that $p_I^{*}\leq D_{max}/2d$ and $p_c^{*}(\beta+a_2)\leq D_{max}/2d$. Thus, $\dfrac{D_{max}}{2d(\alpha+a_1)}-\dfrac{p^{*}_I}{2(\alpha+a_1)}-\dfrac{p^{*}_c(\beta+a_2)}{2(\alpha+a_1)}\geq 0$. Thus, we can write $p_w^{*}$ as 
\begin{eqnarray}
p^{*}_w=\dfrac{D_{max}}{2d(\alpha+a_1)}-\dfrac{p^{*}_I}{2(\alpha+a_1)}-\dfrac{p^{*}_c(\beta+a_2)}{2(\alpha+a_1)}\label{wsppricepull}
\end{eqnarray}
Similarly, we can show that
\begin{eqnarray}
p^{*}_c=\dfrac{D_{max}}{2d(\beta+a_2)}-\dfrac{p^{*}_I}{2(\beta+a_2)}-\dfrac{p^{*}_w(\alpha+a_1)}{2(\beta+a_2)}\label{csppricepull}
\end{eqnarray}
 It is easy to discern that $U_{I,pull}, U_{w,pull}$ and $U_{c,pull}$ are strictly concave in $p_I, p_w$ and $p_c$ respectively , thus, the first order condition is also sufficient. Thus, (\ref{iotspprice})-(\ref{csppricepull}) are optimal. 

We consider the following two cases

{\em Step 3}:
\textit{case i}: We first consider that $\dfrac{D_{max}}{2d}-\dfrac{p^{*}_w(\alpha+a_1)}{2}-\dfrac{p^{*}_c(\beta+a_2)}{2}-\dfrac{ba_1}{2}\geq 0$.
In this case solving for NE strategy profile $(p^{*}_I,p^{*}_c,p_w^{*})$ from (\ref{iotspprice})-(\ref{csppricepull}), we obtain
\begin{eqnarray}
p^{*}_I=\dfrac{D_{max}}{4d}-\dfrac{3ba_1}{4}\label{eq:iotpricepull_lowad}\\
p^{*}_w=\dfrac{D_{max}}{4d(\alpha+a_1)}+\dfrac{ba_1}{4(\alpha+a_1)}\label{eq:wsppricepull_lowad}\\
p^{*}_c=\dfrac{D_{max}}{4d(\beta+a_2)}+\dfrac{ba_1}{4(\beta+a_2)}\label{eq:csppricepull_lowad}
\end{eqnarray}
From (\ref{eq:wsppricepull_lowad}) and (\ref{eq:csppricepull_lowad}) note that this case only arises when $dba_1\leq D_{max}/3$. 

{\em Step 4}:
\textit{case ii}: Now, we consider the case $\dfrac{D_{max}}{2d}-\dfrac{p^{*}_w(\alpha+a_1)}{2}-\dfrac{p^{*}_c(\beta+a_2)}{2}-\dfrac{ba_1}{2}<0$.

From (\ref{iotspprice}),
\begin{align}\label{eq:iotpricepull_highad}
p_I^{*}=0
\end{align}
Solving for $p^{*}_w$ and $p^{*}_c$ with putting $p_I^{*}=0$ we obtain
\begin{align}
p^{*}_w=\dfrac{D_{max}}{3d(\alpha+a_1)}\label{eq:wsppricepull_highad}\\
p^{*}_c=\dfrac{D_{max}}{3d(\beta+a_2)}\label{eq:csppricepull_highad}
\end{align}
Note that 
\begin{align}
\dfrac{D_{max}}{2d}-\dfrac{p_w(\alpha+a_1)}{2}-\dfrac{p_c(\beta+a_2)}{2}-\dfrac{ba_1}{2}\nonumber\\
=\dfrac{D_{max}}{6d}-\dfrac{ba_1}{2}<0 \quad (\text{when }dba_1>D_{max}/3d)\nonumber
\end{align}
Thus, this case arises only when $dba_1>\dfrac{D_{max}}{3d}$. 
Thus from case i and ii, we obtain when $dba_1\leq D_{max}/3$, then the NE price strategy must be of the form given in (\ref{eq:iotpricepull_lowad})-(\ref{eq:csppricepull_lowad}) and when $dba_1>D_{max}/3$, then the NE price strategy is of the form given in (\ref{eq:iotpricepull_highad})-(\ref{eq:csppricepull_highad}). \qed
\textit{Proof of Corollary~\ref{cor:pull_lowad}}
We first obtain the expression of $D_{pull}$ at the equilibrium from Theorem~\ref{thm:pull} when $dba_1\leq D_{max}/3$. Subsequently, we obtain the expression of the value of $U_{I,pull}$, $U_{w,pull}$ and $U_{c,pull}$ at the equilibrium using the already obtained expression of $D_{pull}$ and Theorem~\ref{thm:pull}.

When $dba_1\leq D_{max}/3$, then the equilibrium demand from (\ref{demandpull}), and Theorem~\ref{thm:pull} is 
\begin{align}\label{demandlowad_pull}
D_{pull}& =D_{max}-d(\dfrac{D_{max}}{4d}-\dfrac{3ba_1}{4}+\dfrac{D_{max}}{4d}\nonumber\\& +\dfrac{ba_1}{4}+\dfrac{D_{max}}{4d}+\dfrac{ba_1}{4})\nonumber\\
& =\dfrac{D_{max}}{4}+d\dfrac{ba_1}{4}
\end{align}
Now, we obtain the equilibrium value of $U_{I,pull}$. From (\ref{demandlowad_pull}), (\ref{utilitypull}) and Theorem~\ref{thm:pull}, we obtain
\begin{align}\label{utilitypulliotspfinal}
U_{I,pull}& =(\dfrac{D_{max}}{4d}-\dfrac{3ba_1}{4})(\dfrac{D_{max}}{4}+d\dfrac{ba_1}{4})\nonumber\\
& +ba_1(\dfrac{D_{max}}{4}+d\dfrac{ba_1}{4})\nonumber\\
& =d\left(\dfrac{D_{max}}{4d}+\dfrac{ba_1}{4}\right)^2
\end{align} 
Now, we obtain the expressions of the equilibrium payoffs of the WSP and CSP. From (\ref{utilitypullwsp}),  (\ref{utilitypullcsp}) and Theorem~\ref{thm:pull} we obtain
\begin{eqnarray}
U_{w,pull}& =d(\dfrac{D_{max}}{4d}+\dfrac{ba_1}{4})^2\label{utilitypullwspfinal}\\
U_{c,pull}& =d(\dfrac{D_{max}}{4d}+\dfrac{ba_1}{4})^2\label{utilitypullcspfinal}
\end{eqnarray}
Hence, the result directly follows from (\ref{demandlowad_pull})-(\ref{utilitypullcspfinal}).\qed

\textit{Proof of Corollary~\ref{cor:pull_highad}}
Similar to Corollary~\ref{cor:pull_lowad} we first obtain the expression of the demand at the equilibrium pricing strategy using Theorem~\ref{thm:pull}. Subsequently, we obtain the expressions of the $U_{I,pull}, U_{w,pull}$ and $U_{c,pull}$ using the expression of demand and Theorem~\ref{thm:pull} when $dba_1>D_{max}/3$.

When $dba_1>D_{max}/3$, then by Theorem~\ref{thm:pull}
$p_I^{*}=0$, thus, by Theorem~\ref{thm:pull} and (\ref{demandpull}) we obtain
\begin{eqnarray}\label{demand_pull_highad}
D_{pull}=&D_{max}-d(\dfrac{D_{max}}{3d}+\dfrac{D_{max}}{3d})=\dfrac{D_{max}}{3}
\end{eqnarray}
Now, we obtain the expression of $U_{I,pull}$ for $dba_1>D_{max}/3$. From Theorem~\ref{thm:pull} and (\ref{utilitypull}) we obtain
\begin{align}
U_{I,pull}=& p_I\dfrac{D_{max}}{3}+ba_1\dfrac{D_{max}}{3}=ba_1\dfrac{D_{max}}{3}\label{eq:iotpullhighadpay1}
\end{align}
Now, we obtain the expressions of $U_{w,pull}$ and $U_{c,pull}$ respectively when $dba_1>D_{max}/3$. From (\ref{utilitypullwsp}), (\ref{demand_pull_highad}), and Theorem~\ref{thm:pull}, we obtain
\begin{align}\label{eq:wsppullhighadpay}
U_{w,pull}=& p_w(\alpha+a_1)D_{max}/3=D_{max}^2/9d
\end{align}
From (\ref{utilitypullcsp}), (\ref{demand_pull_highad}), and Theorem~\ref{thm:pull} we obtain
\begin{align}\label{eq:csppullhighadpay}
U_{c,pull}=& p_c(\beta+a_2)D_{max}/3=D_{max}^2/9d
\end{align}
The result readily follows from (\ref{demand_pull_highad})-(\ref{eq:csppullhighadpay}).\qed

\subsection{Proofs of Results in Section~\ref{sec:hybrid}}\label{proof:hybrid}
We first show Lemma~\ref{lem:pwpi} and Theorem~\ref{thm:csp_hybrid}. Using those results we outline the proofs of Corollaries~\ref{cor:hybrid_lowad} and \ref{cor:hybrid_highad}. 

\textit{Proof of Lemma~\ref{lem:pwpi}:}
We first obtain the expressions of the payoffs of the IoTSP and WSP using the expression of $D_{hybrid}$ in (\ref{utilityhybrid}) and (\ref{utilityhybridwsp}) (Step 1). Subsequently, we show that $U_{I,hybrid}$ and $U_{w,hybrid}$ are strictly concave in $p_I$ and $p_w$ respectively, thus, the first order condition is necessary and sufficient (Step 2). We characterize the equilibrium prices $p_I^{*}, p_w^{*}$ for a given $p_c$  in two possible cases depending on whether $p_I^{*} $ is positive or $0$ (Step 3). The detailed proof is given below.

{\em Step 1}: Using (\ref{demandhybrid})   in (\ref{utilityhybrid}) and (\ref{utilityhybridwsp}), we obtain
\begin{align}
U_{I,hybrid}& =(p_I+ba_1-p_c(\beta+a_2))(D_{max}-d(p_I+p_w(\alpha+a_1)))\nonumber\\
U_{w,hybrid}& =p_w(\alpha+a_1)(D_{max}-d(p_I+p_w(\alpha+a_1)))\nonumber
\end{align}
{\em Step 2}: Note that $U_{I,hybrid}$ and $U_{w,hybrid}$ are strictly concave in $p_I$ and $p_w$ respectively. Thus, the first order condition is necessary and sufficient for optimality. Note that IoTSP and WSP simultaneously select price knowing $p_c$. Since $p_I^{*}, p_w^{*}$ are non negative, thus, from the first order condition--
\begin{eqnarray}
p^{*}_I=\max\{\dfrac{D_{max}}{2d}-\dfrac{p_w^{*}(\alpha+a_1)}{2}+\dfrac{p_c(\beta+a_2)}{2}-\dfrac{ba_1}{2},0\}\label{iotpricehybrid}\\
p^{*}_w=\max\{\dfrac{D_{max}}{2d(\alpha+a_1)}-\dfrac{p^{*}_I}{2(\alpha+a_1)},0\}\label{hybridwsp1}
\end{eqnarray}
Note that $p_I^{*}\leq D_{max}/d$, otherwise the demand will be $0$. Thus, 
\begin{align}
p^{*}_w=\dfrac{D_{max}}{2d(\alpha+a_1)}-\dfrac{p^{*}_I}{2(\alpha+a_1)}
\end{align}

{\em Step 3}: We analyze two possible cases:

\textit{Case i}:
$\dfrac{D_{max}}{2d}-\dfrac{p_w^{*}(\alpha+a_1)}{2}+\dfrac{p_c(\beta+a_2)}{2}-\dfrac{ba_1}{2}> 0$.

We obtain from (\ref{iotpricehybrid}) and (\ref{hybridwsp1}) at NE strategy $(p_I^{*},p_w^{*})$:
\begin{eqnarray}
p^{*}_I=\dfrac{D_{max}}{3d}-\dfrac{2ba_1}{3}+\dfrac{2p_c(\beta+a_2)}{3}\nonumber\\
p^{*}_w=\dfrac{D_{max}}{3d(\alpha+a_1)}+\dfrac{ba_1}{3(\alpha+a_1)}-\dfrac{p_c(\beta+a_2)}{3(\alpha+a_1)}\nonumber
\end{eqnarray}
Since $p_I^{*}\leq D_{max}/d$, thus, $p_c(\beta+a_2)\leq \dfrac{D_{max}}{d}+ba_1$. Thus, this case only arises when $p_c(\beta+a_2)>ba_1-\dfrac{D_{max}}{2d}$ and $p_c(\beta+a_2)\leq\dfrac{D_{max}}{d}+ba_1$.

\textit{Case ii}: $\dfrac{D_{max}}{2d}-\dfrac{p_w^{*}(\alpha+a_1)}{2}+\dfrac{p_c(\beta+a_2)}{2}-\dfrac{ba_1}{2}\leq 0$.

By (\ref{iotpricehybrid}), $p_I^{*}=0$, thus, we always have $D_{max}>dp_I^{*}$ and thus, $p_w^{*}>0$ by (\ref{hybridwsp1}) . By (\ref{hybridwsp1}), we also obtain
\begin{align}
p_w^{*}=\dfrac{D_{max}}{2d(\alpha+a_1)}
\end{align}
Hence, the result follows.\qed

\textit{Proof of Theorem~\ref{thm:csp_hybrid}:}
There are two possible  cases of $p_I^{*}$ and $p_w^{*}$ depending on $p_c^{*}$ as given in Lemma~\ref{lem:pwpi}. We investigate each such case, and characterize the equilibrium value of $p_c$ in each case if it exists. 
 
i) We first consider the case where $ p_c(\beta+a_2)> ba_1-\dfrac{D_{max}}{2d}$ (Step 1). We show that the payoff of the CSP is strictly concave function in $p_c$ in this case and thus, the first order condition is both necessary and sufficient for the equilibrium (Step 1a). We characterize the equilibrium in this case (Step 1b).

ii) We next consider the case $p_c(\beta+a_2)\leq ba_1-\dfrac{D_{max}}{2d}$ (Step 2). In this case, the payoff function of the CSP is not strictly concave in $p_c$ (Step 2a). We show that when $p_c(\beta+a_2)<ba_1-\dfrac{D_{max}}{2d}$, then no NE exists  since the CSP always has profitable deviation (Step 2b). However,  we show that when $p_c(\beta+a_2)=ba_1-\dfrac{D_{max}}{2d}$, then the equilibrium exists. We  obtain the necessary and sufficient condition  for existence of the equilibrium (Step 2c).

\textit{Case i}: 
\begin{align}\label{positiveiotprice}
p_c(\beta+a_2)>ba_1-\dfrac{D_{max}}{2d}
\end{align}
{\em Step 1a}:
$p_I^{*}$ and $p_w^{*}$ are given by (\ref{eq:iotsphybrid1}) and (\ref{eq:wsphybrid1}) respectively in Theorem~\ref{thm:hybrid}. Replacing the values of $p^{*}_I$ and $p_w^{*}$ in (\ref{demandhybrid})  the equilibrium demand becomes
\begin{align}\label{demandhybridcsp}
D_{hybrid}
& =\dfrac{D_{max}}{3}+d(\dfrac{ba_1}{3}-\dfrac{p_c(\beta+a_2)}{3})
\end{align}
Hence, CSP\rq{}s payoff from (\ref{utilityhybridcsp}) is 
\begin{align}\label{csppayoffhy}
p_c(\beta+a_2)(\dfrac{D_{max}}{3}+d\dfrac{ba_1}{3}-d\dfrac{p_c(\beta+a_2)}{3})
\end{align}
{\em Step 1b}:
From first order condition, we obtain
\begin{align}\label{csphybridcasei}
p_c^{*}=\dfrac{D_{max}}{2d(\beta+a_2)}+\dfrac{ba_1}{2(\beta+a_2)}
\end{align}
which is optimal by the strict concavity of (\ref{csppayoffhy}). 

Condition in (\ref{positiveiotprice}) is satisfied when $\dfrac{D_{max}}{2d}+\dfrac{ba_1}{2}>ba_1-\dfrac{D_{max}}{2d}$ i.e. $dba_1<2D_{max}$.

\textit{case ii}: $p_c(\beta+a_2)\leq ba_1-\dfrac{D_{max}}{2d}$:

{\em Step 2a}:
In this case $p^{*}_I=0$ and $p_w^{*}(\alpha+a_1)=\dfrac{D_{max}}{2d}$ by (\ref{eq:iotsphybrid3}).  
Hence, from (\ref{demandhybrid}) the equilibrium demand  is 
\begin{align}\label{eq:demandlowad}
D_{hybrid}=D_{max}/2
\end{align}
Thus, CSP\rq{}s payoff is 
\begin{align}
p_c(\beta+a_2)D_{max}/2\nonumber
\end{align}
which is strictly increasing in $p_c$. 

{\em Step 2b}: If $p_c(\beta+a_2)< ba_1-\dfrac{D_{max}}{2d}$ is an NE price, then we can find a small enough $\epsilon$ such that $(p_c+\epsilon)(\beta+a_2)<ba_1-\dfrac{D_{max}}{2d}$, thus at $p_c+\epsilon$ the payoff is $(p_c+\epsilon)(\beta+a_2)D_{max}/2$ which is strictly larger compared to the payoff at $p_c$ contradicting the fact that $p_c$ is an NE. Thus, there is no NE at $p_c(\beta+a_2)<ba_1-\dfrac{D_{max}}{2d}$.



{\em Step 2c}: Now, we investigate the setting when
\begin{align}\label{cspopti}
p_c(\beta+a_2)=ba_1-\dfrac{D_{max}}{2d}
\end{align}
Thus, the CSP\rq{}s payoff is now
\begin{align}\label{eq:csphybridhighad}
(ba_1-\dfrac{D_{max}}{2d})D_{max}/2
\end{align}
Now, for an NE strategy we have to rule out any profitable unilateral deviation by CSP. Toward this end we consider the two following cases:

\textit{Case ii. a}: First we rule out the possibility that the CSP selects a price $x$ which is lower than $p_c$ (given in (\ref{cspopti})). 

Note that if $x(\beta+a_2)<ba_1-\dfrac{D_{max}}{2d}$, then the optimal demand is $D_{max}/2$. Hence, CSP\rq{}s payoff is $x(\beta+a_2)D_{max}/2$ which is strictly less than (\ref{eq:csphybridhighad}) since $x(\beta+a_2)<ba_1-\dfrac{D_{max}}{2d}$. Thus, CSP does not have any incentive to select a price lower than that of (\ref{cspopti}).

\textit{Case ii.b}: Now, we obtain the condition under which CSP will not have any incentive to select a higher price compared to $p_c$ (given in (\ref{cspopti})). 

If CSP selects price $x>p_c$, then $x(\beta+a_2)>ba_1-\dfrac{D_{max}}{2d}$ and the condition in case i is satisfied, thus, IoTSP and WSP will select the price according to (\ref{eq:iotsphybrid1}) and (\ref{eq:wsphybrid1}) respectively with $x$ in place of $p_c$. Thus, CSP\rq{}s payoff becomes (from (\ref{csppayoffhy}))
\begin{align}
x(\beta+a_2)(\dfrac{D_{max}}{2}-\dfrac{D_{max}}{6}+dba_1/3-dx(\beta+a_2)/3)\nonumber
\end{align}
For no profitable unilateral deviation, we must have from (\ref{eq:csphybridhighad})
\begin{align}
& (ba_1-\dfrac{D_{max}}{2d}-x(\beta+a_2))(\dfrac{D_{max}}{2}-dx(\beta+a_2)/3)\geq 0\nonumber
\end{align}
Since $x(\beta+a_2)>ba_1-\dfrac{D_{max}}{2d}$ thus the above condition is satisfied only when $3\dfrac{D_{max}}{2d}\leq x(\beta+a_2)$. If $ ba_1-\dfrac{D_{max}}{2d}< 3\dfrac{D_{max}}{2d}$, then we can find a $x$ such that $3\dfrac{D_{max}}{2d}>x(\beta+a_2)>ba_1-\dfrac{D_{max}}{2d}$. Thus, we must have $3\dfrac{D_{max}}{2d}\leq ba_1-\dfrac{D_{max}}{2d}$ or $dba_1\geq 2D_{max}$. Thus, the above condition can only hold when $dba_1\geq 2D_{max}$.

Thus from case i and ii, when $dba_1<2D_{max}$, then $p_c^{*}$ is given by (\ref{csphybridcasei}); on the other hand, when $dba_1\geq 2D_{max}$, then the NE strategy profile must be of the form given in (\ref{cspopti}).  Hence, the result follows. \qed

\textit{Proof of Corollary~\ref{cor:hybrid_lowad}}
First, we obtain the expression of the demand at the equilibrium pricing strategies using Theorem~\ref{thm:hybrid}. Subsequently, we obtain the expressions of the payoffs of the IoTSP, WSP and CSP at the equilibrium when $dba_1<2D_{max}$ using Theorems~\ref{thm:csp_hybrid}, \ref{thm:hybrid} and the expression of $D_{hybrid}$.

From (\ref{demandhybrid}), and Theorem~\ref{thm:hybrid} the equilibrium demand is
\begin{align}\label{demandhybrid_lowad}
D_{hybrid}& =D_{max}-d(p^{*}_I+p^{*}_w(\alpha+a_1))\nonumber\\
& =D_{max}-d(\dfrac{2D_{max}}{3d}-ba_1/3+\dfrac{ba_1}{6}+\dfrac{D_{max}}{6d})\nonumber\\
& =\dfrac{D_{max}}{6}+d\dfrac{ba_1}{6}
\end{align}
 Now, we obtain the expression of $U_{I,hybrid}$ and $U_{w,hybrid}$ at the equilibrium. From (\ref{utilityhybrid}), (\ref{demandhybrid_lowad}), and Theorem~\ref{thm:hybrid} we obtain
\begin{align}
U_{I,hybrid}& =(\dfrac{2D_{max}}{3d}-\dfrac{ba_1}{3})(\dfrac{D_{max}}{6}+d\dfrac{ba_1}{6})\nonumber\\& 
+ba_1(\dfrac{D_{max}}{6}+d\dfrac{ba_1}{6})\nonumber\\& -(\dfrac{D_{max}}{2d}+\dfrac{ba_1}{2})(\dfrac{D_{max}}{6}+d\dfrac{ba_1}{6})\nonumber\\
& =d\left(\dfrac{D_{max}}{6d}+\dfrac{ba_1}{6}\right)^2\label{eq:utilityiothybridlowad}\\
U_{w,hybrid}& =d(\dfrac{D_{max}}{6d}+\dfrac{ba_1}{6})^2\label{eq:hybridwspfinal}
\end{align}
%
Now, we obtain the expression of $U_{c,hybrid}$ at the equilibrium. From (\ref{demandhybrid_lowad}) and (\ref{thm:csp_hybrid}) we obtain
\begin{eqnarray}\label{eq:hybridcspfinal}
U_{c,hybrid}& =3d(\dfrac{D_{max}}{6d}+\dfrac{ba_1}{6})^2.
\end{eqnarray}
The result follows from (\ref{demandhybrid_lowad})-(\ref{eq:hybridcspfinal}).\qed

\textit{Proof of Corollary~\ref{cor:hybrid_highad}}
Similar to Corollary~\ref{cor:hybrid_highad} we obtain the expression of the demand at the equilibrium using Theorem~\ref{thm:hybrid} when $dba_1\geq 2D_{max}$. Subsequently, we obtain the expressions of $U_{I,hybrid}, U_{w,hybrid}$ and $U_{c,hybrid}$ when $dba_1\geq 2D_{max}$ using Theorems~\ref{thm:csp_hybrid} and \ref{thm:hybrid}. 

From (\ref{demandhybrid}) and Theorem~\ref{thm:hybrid} we obtain
\begin{align}\label{demandhighadhybrid}
D_{hybrid}& =D_{max}-d(p_I^{*}+p_w^{*}(\alpha+a_1))\nonumber\\
& =\dfrac{D_{max}}{2}
\end{align}
Now, we obtain the expression of $U_{I,hybrid}$. From (\ref{utilityhybrid}) and Theorems~\ref{thm:csp_hybrid},\ref{thm:hybrid} we obtain 
\begin{align}\label{utilityiotsphybridhighad}
U_{I,hybrid}=(ba_1-p_c(\beta+a_2))\dfrac{D_{max}}{2}=\dfrac{D_{max}^{2}}{4d}
\end{align}
Also note from Theorem~\ref{thm:hybrid}, (\ref{demandhighadhybrid}) and (\ref{utilityhybridwsp}) that
\begin{align}\label{utilitywsphybridhighad}
U_{w,hybrid}=\dfrac{D_{max}^2}{4d}
\end{align}
Now, we obtain the expression of $U_{c,hybrid}$. From Theorem~\ref{thm:csp_hybrid} and (\ref{demandhighadhybrid}) we obtain
\begin{align}\label{payoffcsphybridfinal}
U_{c,hybrid}& =(ba_1-\dfrac{D_{max}}{2d})D_{max}/2
\end{align}
The result readily follows from (\ref{demandhighadhybrid})-(\ref{payoffcsphybridfinal}).\qed

\subsection{Proofs of Results in Section~\ref{sec:comparison_models}}\label{proof:compare}
We provide the outlines of the proofs of Corollaries~\ref{cor:demand_compare},\ref{cor:iot_compare}, \ref{cor:wsp_compare} and \ref{cor:csp_compare}. 

\textit{Proof of Corollary~\ref{cor:demand_compare}}
First we compare the demand of end-users in different models when $dba_1\leq D_{max}/3$ (Step 1). Subsequently, we compare the demand of end-users in different models when $2D_{max}\leq dba_1>D_{max}/3$ (Step 2) and $5D_{max}>dba_1>2D_{max}$ (Step 3). Finally, we consider the case when $dba_1\geq 5D_{max}$ (Step 4). This will complete the proof.

{\em Step 1}: When $dba_1\leq D_{max}/3$, then from Corollaries~\ref{cor:push_lowad} and \ref{cor:hybrid_lowad}, $D_{push}=D_{hybrid}$. Also from Corollary~\ref{cor:pull_lowad} $D_{pull}>D_{push}$. 

 {\em Step 2}:When $2D_{max}\leq dba_1>D_{max}/3$, then from Corollaries~\ref{cor:push_lowad} and \ref{cor:hybrid_lowad} $D_{push}=D_{hybrid}$. But $D_{pull}$ remains constant at $D_{max}/3$. Thus, 
\begin{align}
& D_{hybrid}=D_{push}>D_{pull}\quad if 2D_{max}\leq dba_1>D_{max}\nonumber\\
& D_{hybrid}=D_{push}=D_{pull}\quad if dba_1=D_{max}\nonumber\\
& D_{hybrid}=D_{push}<D_{pull}\quad if D_{max}/3<dba_1<D_{max}\nonumber
\end{align}

{\em Step 3}:When $5D_{max}>dba_1>2D_{max}$, then $D_{push}>D_{hybrid}$ from Corollaries~\ref{cor:push_lowad} and \ref{cor:hybrid_highad}. Also note from Corollary~\ref{cor:pull_highad} $D_{push}>D_{pull}$ and $D_{hybrid}>D_{pull}$.

{\em Step 4}:When $dba_1\geq 5D_{max}$, then $D_{push}=D_{max}$ by Corollary~\ref{cor:push_highad} which is higher than $D_{hybrid}$ (Corollary~\ref{cor:hybrid_highad}) and $D_{pull}$ (Corollary~\ref{cor:pull_highad}).
 Hence, the  result follows.\qed

\textit{Proof of Corollary~\ref{cor:iot_compare}}
We first compare the payoff of the IoTSP in the push and pull models (Step 1). Towards this end we first show that when $dba_1\leq D_{max}/3$, then $U_{I,push}<U_{I,pull}$ using Corollaries~\ref{cor:push_lowad} and \ref{cor:pull_lowad} (Step 1a). Subsequently, we show that that $U_{I,push}-U_{I,pull}$ is strictly decreasing when $D_{max}/3\leq dba_1<5D_{max}$ using Corollaries~\ref{cor:pull_highad} and \ref{cor:push_lowad}. Since when $dba_1=D_{max}/3$, $U_{I,pull}>U_{I,push}$, thus,  it readily follows that $U_{I,push}<U_{I,pull}$ when $dba_1<5D_{max}$ (Step 1b). Finally, we show that even when $dba_1\geq 5D_{max}$, then $U_{I,push}<U_{I,pull}$ using Corollaries~\ref{cor:pull_highad} and \ref{cor:push_highad} (Step 2). Finally, we show that $U_{I,hybrid}=U_{I,push}$ when $dba_1\leq 2D_{max}$, however, $U_{I,push}>U_{I,hybrid}$ when $dba_1>2D_{max}$ (Step 3). This will complete the proof. The detailed proof is given below.

{\em Step 1}: We first compare between $U_{I,push}$ and $U_{I,pull}$.

{\em Step 1a}: $dba_1\leq D_{max}/3$

Comparing Corollaries~\ref{cor:push_lowad}, and \ref{cor:pull_lowad} we obtain $U_{I,push}<U_{I,pull}$ in this case.

 {\em Step 1b}: $D_{max}/3<dba_1\leq 5D_{max}$
 
 From Corollaries~\ref{cor:push_lowad} and \ref{cor:pull_highad} the difference between  $U_{I,push}$ and $U_{I,pull}$ is 
\begin{align}
d(\dfrac{D_{max}}{6d}+\dfrac{ba_1}{6})^2-ba_1D_{max}/3\nonumber
\end{align}
It is easy to verify that the above expression is strictly decreasing in $ba_1$ for $dba_1<5D_{max}$. But when $dba_1=D_{max}/3$, the above expression is negative. Thus, $U_{I,push}<U_{I,pull}$ for $5D_{max}>dba_1>D_{max}/3$. 

{\em Step 1c}: $dba_1\geq 5D_{max}$. 

Note from Corollary~\ref{cor:push_highad} that when $dba_1\geq 5D_{max}$, then $U_{I,push}=D_{max}^2/d$, but from Corollary~\ref{cor:pull_highad} $U_{I,pull}$ is $ba_1D_{max}/3$ which is strictly higher compared to $D_{max}^2/d$ for $dba_1\geq 5D_{max}$. 

Thus, $U_{I,push}<U_{I,pull}$ for all sets of parameters. 

{\em Step 2}: Now, we compare the payoffs of the IoTSP in the push and hybrid models. 

When $2D_{max}> dba_1$, then from Corollaries~\ref{cor:push_lowad} and \ref{cor:hybrid_lowad} we obtain $U_{I,push}=U_{I,hybrid}$. 

When $dba_1>2D_{max}$ $U_{I,push}>\dfrac{D_{max}^2}{4d}$ (from Corollaries~\ref{cor:hybrid_highad}  and \ref{cor:push_lowad}, \ref{cor:push_highad}), thus, $U_{I,push}>U_{I,hybrid}$. 

Since $U_{I,hybrid}$ never exceeds $U_{I,push}$ and $U_{I,push}<U_{I,pull}$. Thus, $U_{I,hybrid}<U_{I,pull}$. Hence, the result follows.\qed

\textit{Proof of Corollary~\ref{cor:wsp_compare}} 
First we compare the payoffs of the WSP in the push and pull models (Step 1). Towards this end, we show that when $dba_1\leq D_{max}/3$, $U_{w,pull}>U_{w,push}$ from Corollaries~\ref{cor:push_lowad} and \ref{cor:pull_lowadad}(Step 1.a). However, we show that $U_{w,push}-U_{w,pull}$ increases with $ba_1$ $D_{max}/(3d)<dba_1<5D_{max}$. $U_{w,push}<U_{w,pull}$ when $dba_1<0.414D_{max}$. However, $U_{w,push}$ exceeds $U_{w,pull}$ when $dba_1>0.414 D_{max}$ (Step 1.b). Finally, we show that even the worst case payoff of the WSP $\tilde{U}_{w,push}$ is higher than $U_{w,pull}$ when $dba_1\geq 5D_{max}$ (Step 1.c.). Thus, $U_{w,push}<U_{w,pull}$ when $dba_1<0.414D_{max}$ and $U_{w,pull}<U_{w,push}$ when $dba_1>0.414D_{max}$.

Proceeding as similar as above Subsequently, we compare the payoffs of the WSP in the push and hybrid models (Step 2) and the pull and hybrid models (Step 3). 

{\em Step 1}: We first compare between $U_{w,push}$ and $U_{w,pull}$.   

{\em Step 1.a.}:$dba_1\leq D_{max}/3$:

 From Corollaries~\ref{cor:push_lowad} and \ref{cor:pull_lowad}  it is easy to discern that $U_{w,push}<U_{w,pull}$ when $dba_1\leq D_{max}/3$. 


{\em Step 1.b.}: $5D_{max}>dba_1>D_{max}/3$:

Note from Corollaries~\ref{cor:push_lowad} and \ref{cor:pull_highad}  that when $5D_{max}>dba_1>D_{max}/3$ the difference between WSP\rq{}s payoff in the push model and that of in the pull model is 
\begin{align}
2d\left(\dfrac{D_{max}}{6d}+\dfrac{ba_1}{6}\right)^2-\dfrac{D_{max}^2}{9d}\nonumber
\end{align}
The above expression monotonically increases with $ba_1$. The two payoffs are equal when $ba_1=6(\dfrac{D_{max}}{\sqrt{18}d}-\dfrac{D_{max}}{6d})=0.414D_{max}/d$, after that WSP\rq{}s payoff is higher in the push model compared to the pull model. 

{\em Step 1.c.}:$dba_1>5D_{max}$:

On the other hand note from Corollary~\ref{cor:push_highad} that the worst case WSP\rq{}s payoff in the push model is $2D_{max}^2/d$ when $dba_1\geq 5D_{max}$ which is strictly higher compared to $D_{max}^2/9d$ which is the payoff obtained by WSP in the pull model when $dba_1\geq 5D_{max}$ (by Corollary~\ref{cor:pull_highad}). 

Thus, WSP\rq{}s payoff is higher in the pull model when $dba_1<0.414D_{max}$ and higher in the push model when $dba_1>0.414D_{max}$. Payoffs of WSP in the push model and in the pull model are equal when $dba_1=0.414D_{max}$.

{\em Step 2}: Now, we compare the payoffs in the push and hybrid models.

  From Corollaries~\ref{cor:push_lowad} and \ref{cor:hybrid_lowad} it is easy to discern that $U_{w,push}>U_{w,hybrid}$ when $dba_1<2D_{max}$. 
  
  Also note from Corollary~\ref{cor:push_lowad} and \ref{cor:hybrid_highad}
that $U_{w,push}>U_{w,hybrid}$ when $5D_{max}>dba_1\geq 2D_{max}$. 

When $dba_1\geq 5D_{max}$, then $\tilde{U}_{w,push}$ (the worst case payoff) is strictly higher compared to the payoff in the hybrid model (Corollary~\ref{cor:hybrid_highad}). Thus, $U_{w,push}>U_{w,hybrid}$ for all sets of parameters.

{\em Step 3}: Now, we compare $U_{w,pull}$ and $U_{w,hybrid}$. 

{\em Step 3a}: $dba_1\leq D_{max}/3$:

From Corollaries~\ref{cor:pull_lowad} and \ref{cor:hybrid_lowad} $U_{w,pull}>U_{w,hybrid}$ when $dba_1\leq D_{max}/3$. 

{\em Step 3b}: $2D_{max}>dba_1>D_{max}/3$:

Note from Corollaries~\ref{cor:hybrid_lowad} and \ref{cor:pull_highad}  that when $2D_{max}>dba_1>D_{max}/3$ the difference between WSP\rq{}s payoff in the hybrid model and that of in the pull model is 
\begin{align}
d\left(\dfrac{D_{max}}{6d}+\dfrac{ba_1}{6}\right)^2-\dfrac{D_{max}^2}{9d}\nonumber
\end{align}
The above expression monotonically increases with $ba_1$. The two payoffs are equal when $ba_1=D_{max}/d$, after that WSP\rq{}s payoff is higher in the hybrid model compared to the pull model. 

{\em Step 3c}: $dba_1\geq 2D_{max}$:

When $dba_1\geq 2D_{max}$, then by Corollaries~\ref{cor:hybrid_highad} and \ref{cor:pull_highad} $U_{w,hybrid}>U_{w,pull}$. Thus, when $dba_1<D_{max}$, $U_{w,pull}>U_{w,hybrid}$. When $dba_1>D_{max}$, $U_{w,pull}<U_{w,hybrid}$. When $dba_1=D_{max}$, $U_{w,hybrid}=U_{w,pull}$.

Hence, the result follows.\qed

\textit{Proof of Corollary~\ref{cor:csp_compare}}:
The comparison of the payoffs of CSP in the push model and pull model is exactly the same as for the WSP. Thus, we omit it here. We thus, start with comparing the payoffs of the CSP in the push and hybrid model (Step 1). We first show that $U_{c,hybrid}<U_{c,push}$ when $dba_1<5D_{max}$ (Step 1a). When $dba_1\geq 5D_{max}$, the payoff of the CSP in the push model is not unique since the equilibrium is not unique (by Theorem~\ref{thm:push}). We show that the worst case payoff of the CSP is strictly less than $U_{c,hybrid}$ (Step 1b). However, the best possible payoff of the CSP is higher than $U_{c,hybrid}$ when $dba_1>5.5D_{max}$ (Step 1c). Finally, we compare the payoffs of the CSP in the hybrid and pull models and we show that $U_{c,hybrid}>U_{c,pull}$ for all values of the parameters (Step 2).

{\em Step 1}:We compare the payoff of CSP in the push model and the hybrid model. 

{\em Step 1a}: $dba_1< 5D_{max}$

From Corollaries~\ref{cor:push_lowad} and \ref{cor:hybrid_lowad} we obtain $U_{c,hybrid}>U_{c,push}$ for $dba_1<2D_{max}$. 

On then other hand when $5D_{max}>dba_1\geq 2D_{max}$, then the difference between $U_{c,push}$ (Corollary~\ref{cor:push_lowad}) and $U_{c,hybrid}$ (Corollary~\ref{cor:hybrid_highad}) is 
\begin{align}
2d(\dfrac{D_{max}}{6d}+ba_1/6)^2-(ba_1-\dfrac{D_{max}}{2d})D_{max}/2
\end{align}
The above expression is strictly decreasing in $ba_1$ for $ba_1<3.5D_{max}$ and strictly increasing in $ba_1$ for $ba_1>3.5D_{max}$. It attains the minimum value at $ba_1=3.5D_{max}$. Note that both at $dba_1=2D_{max}$ and $dba_1=5D_{max}$, the expression is negative. Thus, $U_{c,hybrid}>U_{c,push}$ when $5D_{max}>dba_1\geq 2D_{max}$. 

{\em Step 1b}: We now compare the worst possible payoff of the CSP ($\tilde{U}_{c,push}$) in the push model and the payoff of the CSP in the hybrid model when $dba_1\geq 5D_{max}.$

When $dba_1\geq 5D_{max}$, the worst case payoff of CSP $\tilde{U}_{c,push}$ in the push model is $2D_{max}^2/d$ (see Corollary~\ref{cor:push_highad}). It is easy discern from Corollary~\ref{cor:hybrid_highad} that $U_{c,hybrid}>\tilde{U}_{c,push}$ when $dba_1\geq 5D_{max}$.

{\em Step 1c}: We now compare the best possible payoff of the CSP in the push model ($U_{c,best,push}$) and the payoff of the CSP in the hybrid model when $dba_1\geq 5D_{max}.$ 

The best possible payoff of the CSP in the push model is $(ba_1-3D_{max}/d)D_{max}$ when $dba_1\geq 5D_{max} $(see Corollary~\ref{cor:push_highad}). Now, $U_{c,best,push}$ (the best possible payoff of the CSP in the push model) $- U_{c,hybrid}$ when $dba_1>5D_{max}$ is
\begin{align}
(ba_1-3D_{max}/d)D_{max}-(ba_1-D_{max}/2d)D_{max}/2
\end{align}
The above expression is strictly positive when $dba_1>5.5D_{max}$, when $dba_1<5.5D_{max}$ it is negative. Thus, $U_{c,best,push}>U_{c,hybrid}$ when $dba_1>5.5D_{max}$ and $U_{c,hybrid}>U_{c,best,push}$ when $dba_1<5.5D_{max}$, $U_{c,hybrid}=U_{c,best,push}$ when $dba_1=5.5D_{max}$.

{\em Step 2}: Now, we compare the payoffs in the hybrid model and the pull model.
 
 Note from Corollaries~\ref{cor:pull_lowad} and \ref{cor:hybrid_lowad} when $dba_1\leq D_{max}/3$, then $U_{c,hybrid}>U_{c,pull}$. 
 
 It is also easy to discern from Corollaries~\ref{cor:hybrid_lowad} and \ref{cor:pull_highad} that when $2D_{max}>dba_1>D_{max}/3$, then the payoff of CSP in the hybrid model is higher compared to the payoff of CSP in the pull model, since $3d(\dfrac{D_{max}}{6d}+ba_1/6)^2>\dfrac{D_{max}^2}{9d}$ when $dba_1>D_{max}/3$. 
 
 When $dba_1\geq 2D_{max}$, then $U_{c,hybrid}$ (Corollary~\ref{cor:hybrid_highad}) is $(ba_1-D_{max}/2d)D_{max}/2$ which strictly higher than $D_{max}^2/(9d)$ for $dba_1\geq 2D_{max}/d$.  
 
 Thus, $U_{c,hybrid}>U_{c,pull}$ for any value of $dba_1$. Hence, the result follows. \qed

\end{document}